\documentclass[article,twocolumn]{revtex4-2}
\usepackage{graphicx} % Required for inserting images
\usepackage[utf8]{inputenc}
\usepackage{amsmath}
\usepackage{amssymb}
\usepackage{amsthm}
\usepackage{mathtools}
\usepackage{physics}
\usepackage[dvipsnames]{xcolor}
\usepackage{bbold}
\usepackage{wrapfig}
\usepackage{caption}
\usepackage{subcaption}
\usepackage{dsfont}
\usepackage{centernot}
\usepackage[margin=1.2in]{geometry}
\usepackage{tabularx}
\usepackage{svg}
\usepackage{comment}
\usepackage[english]{babel}
\usepackage{amsthm}
\theoremstyle{definition}

\usepackage{tikz}
\usepackage{hyperref}
\hypersetup{
    colorlinks,
    citecolor=black,
    filecolor=black,
    linkcolor=black,
    urlcolor=black
}
\usepackage{dcolumn}% Align table columns on decimal point
\usepackage{bm}% bold math
\usepackage{dsfont}
\usepackage{verbatim}
\usepackage{array}
\newcolumntype{L}[1]{>{\raggedright\arraybackslash}p{#1}}  % Left-justified

\newcommand{\proj}[1]{|#1\rangle\langle#1|}
\newcommand{\ox}{\otimes}

\begin{document}

\title{Emergence of Classicality in Wigner's Friend Scenarios}

\author{Tom Rivlin}
\email{tom.rivlin@tuwien.ac.at}
\affiliation{Atominstitut, TU Wien, 1020 Vienna, Austria}

\author{Sophie Engineer}
\affiliation{Institute of Photonics and Quantum Sciences (IPAQS), Heriot-Watt University, Edinburgh, United Kingdom}

\author{Veronika Baumann}
\email{veronika.baumann@oeaw.ac.at}
\affiliation{Institute for Quantum Optics and Quantum Information - IQOQI Vienna,\\1090 Vienna, Austria}
\affiliation{Atominstitut, TU Wien, 1020 Vienna, Austria}

\begin{abstract}

The Wigner's Friend (WF) thought experiment concerns quantum measurements by a `superobserver' of an observer measuring a quantum system. Variations on the setup and its extended versions have seen a resurgence in recent years, in light of a series of no-go theorems that reveal new quantum effects and question the existence of absolute events. But most theoretical and experimental studies of WF scenarios have restricted themselves to a `friend' composed of a single qubit with idealised measurement settings in an idealised lab. In this work, we consider a specific, unitary model of the interaction between the Friend and the system in the presence of a decohering environment. In particular, we study WF scenarios from the perspective of quantum Darwinism (QD). The QD framework is well-suited to studying the questions of observations and agents in quantum theory that WF scenarios raise, as it is concerned with how observers record objective information about a system with access only to its surroundings. Here we describe how to add environments to simple and extended WF scenarios in the QD framework, and present numerical results that study the emergence of classicality, in the form of the Friend's measurement result becoming more objective. In both the simple and extended cases, we also find that the model and the environment obfuscate genuine WF effects and introduce strong restrictions on them. However, we also find a novel form of WF effect that exploits coherence between the Friend and the environment.
\end{abstract}

\maketitle

\section{Introduction}
\label{sec:Introduction}

\noindent The well-known Wigner's Friend paradox~\cite{wigner1995remarks,deutsch1985quantum} is concerned with observations of observers in quantum theory. In its original formulation it features two agents: an \emph{observer} -- Wigner's Friend -- who observes a quantum system, and a so called \emph{superobserver} -- Wigner -- who observes both the system and his Friend. The key point is that to Wigner, his Friend is another quantum system over which he is said to have \emph{full quantum control}, despite the system potentially being highly complex. The thought experiment is set up such that Wigner and his Friend will have different accounts of the Friend's interaction with the quantum system she observes. To the Friend this interaction constitutes a measurement that collapses the state. To Wigner, however, provided that the Friend and system are sufficiently isolated
(usually the Friend and the system together are referred to as the Friend's Lab), the same interaction should be described by unitary evolution entangling the  Friend with the system she observes.
In recent years, extended Wigner's Friend scenarios (EWFS), which combine Wigner's Friend experiments with non-locality arguments, have been extensively studied~\cite{brukner2018no,18FrauchigerRenner,20BongUtrerasAlaconGhafari,proietti2019experimental}.
For an EWFS where two halves of an entangled pair of particles are sent to two Wigner's Friend setups, one can derive so called local Friendliness (LF) inequalities and show that they can be violated~\cite{20BongUtrerasAlaconGhafari}. This is significant as it leads to difficult-to-reconcile logical inconsistencies and is often said to be a stronger non-classical effect than the failure of local hidden variables to explain Bell scenarios.

In this work, we study a modified, more complex description of Wigner's Friend scenarios, where the Lab also contains an environment that causes decoherence on the quantum system~\cite{zurek2003decoherence,schlosshauer2019quantum}. A similar modification was made in~\cite{20Relano}, and we will highlight where our approach compares and differs with the one from that work. The purpose of our model is to understand whether and to what extent Wigner's Friend effects are still possible when decoherence occurs, and we do indeed find WF-type effects
in the presence of a decohering environment.
More concretely, we focus on an extension of decoherence theory called quantum Darwinism (QD)~\cite{09Zurek,21Korbicz,22Zurek}. This theory examines in detail how information flows between a system and its environment during the decoherence process. In QD, the environment acts as a `witness' of the information contained within a system, meaning that an observer with access to fractions of the environment would be able to discern information about the system. Key to this process is that only certain kinds of information -- the information contained in the \textit{pointer basis} of the system -- can `survive' the broadcasting process and be easily accessible from the environment. In the case of perfect broadcasting, this pointer basis information constitutes \emph{objective classical information}.

In a recent series of works~\cite{23SchwarzhansBinderHuber, engineer2024equilibration}, the QD framework was used as the basis for a unitary, dynamical model of the measurement process. In particular, the measurement was modelled as a \textit{closed-system equilibration} process commensurate with principles of statistical mechanics. This was motivated by works that questioned the thermodynamic consistency of conventional treatments of measurements in quantum theory~\cite{19ManzanoGuryanovaHuber,20GuryanovaFriisHuber}. Inspired by this, here we seek a model of WF scenarios that, in principle, can extend to treating the Friend as a macroscopic system (see~\cite{21ZukowskiMarkiewicz,de2024finite} for other recent works that have considered thermodynamic implications of Wigner's Friend).

In other words, given that the Friend may be a macroscopic object, or indeed a person, here we are interested in asking \emph{what precisely happens inside the lab when the Friend makes her measurement?} Here we model the Friend's measurement as a unitary decoherence process that broadcasts information from the system into the environment. We then consider the Friend to be one so-called \emph{macrofraction} of the environment -- a large grouping of environment parts. One goal of this work is to seek some form of \emph{emergence of classicality} as the Friend increases in size. In QD this can manifest as the emergence of a feature called \emph{objectivity}.

A benefit of our model is that it allows us to treat the Friend as a quantum system with increasing complexity. There has been much debate in recent years as to what constitutes an `observer', or an `agent' in quantum theory. Is a simple, single qubit enough, or does the Friend need to be able to reason about the laws of quantum theory in some way? \cite{nurgalieva2018inadequacy,nurgalieva2022thought,23WisemanCavalcantiRieffel,24WalleghamYingWagner,zeng2024towards} (Indeed this is a debate Wigner himself considered~\cite{95Wigner,99Esfield}.) Our model allows us to study the properties of the quantum system we call the Friend, as a function of her Hilbert space dimension. Later we will show numerical results for the magnitude of the WF-type effects as a function of how many qubits the Friend is composed of, and show that classicality (in the form of the vanishing of these effects) can be seen to emerge as the Friend increase in size.

The paper is structured as follows. In Sec.~\ref{sec:Wigner's Friend} we consider the `simple' Wigner's Friend experiment consisting of one quantum system, one Friend and one Wigner. We start by reviewing (a modern version of) the original thought experiment in Sec.~\ref{ssec:The simple Wigner's Friend experiment1}, followed by a version containing an environment in Sec.~\ref{ssec:WFwenv}. We then consider in Sec.~\ref{ssec:WFandQD} a WF scenario where the Friend's measurement is modelled as a QD-based decoherence process.  In Sec.~\ref{ssec:Numerical results1} we present the numerical results pertaining to this model. Sec.~\ref{sec:Extended Wigner's Friend scenarios1} is about extended Wigner's Friend scenarios, and in particular Local Friendliness inequalities. Again, we start by reviewing a standard EWFS and the LF no-go theorem in Sec.~\ref{ssec:LF}, and then apply our QD-based model to these setups in Sec.~\ref{ssec:Extended Wigner's Friend experiments2}, discussing its implications for violations of LF inequalities. This is then followed by a presentation of further numerical results in Sec.~\ref{ssec:Numerical results2}.
Finally, our conclusions are summarized in Sec.~\ref{sec:Conclusions and Outlook}.

\section{The Wigner's Friend experiment}
\label{sec:Wigner's Friend}

\subsection{The simple Wigner's Friend scenario}
\label{ssec:The simple Wigner's Friend experiment1}

\begin{figure*}[htbp]
    \includegraphics[width=0.45\linewidth]{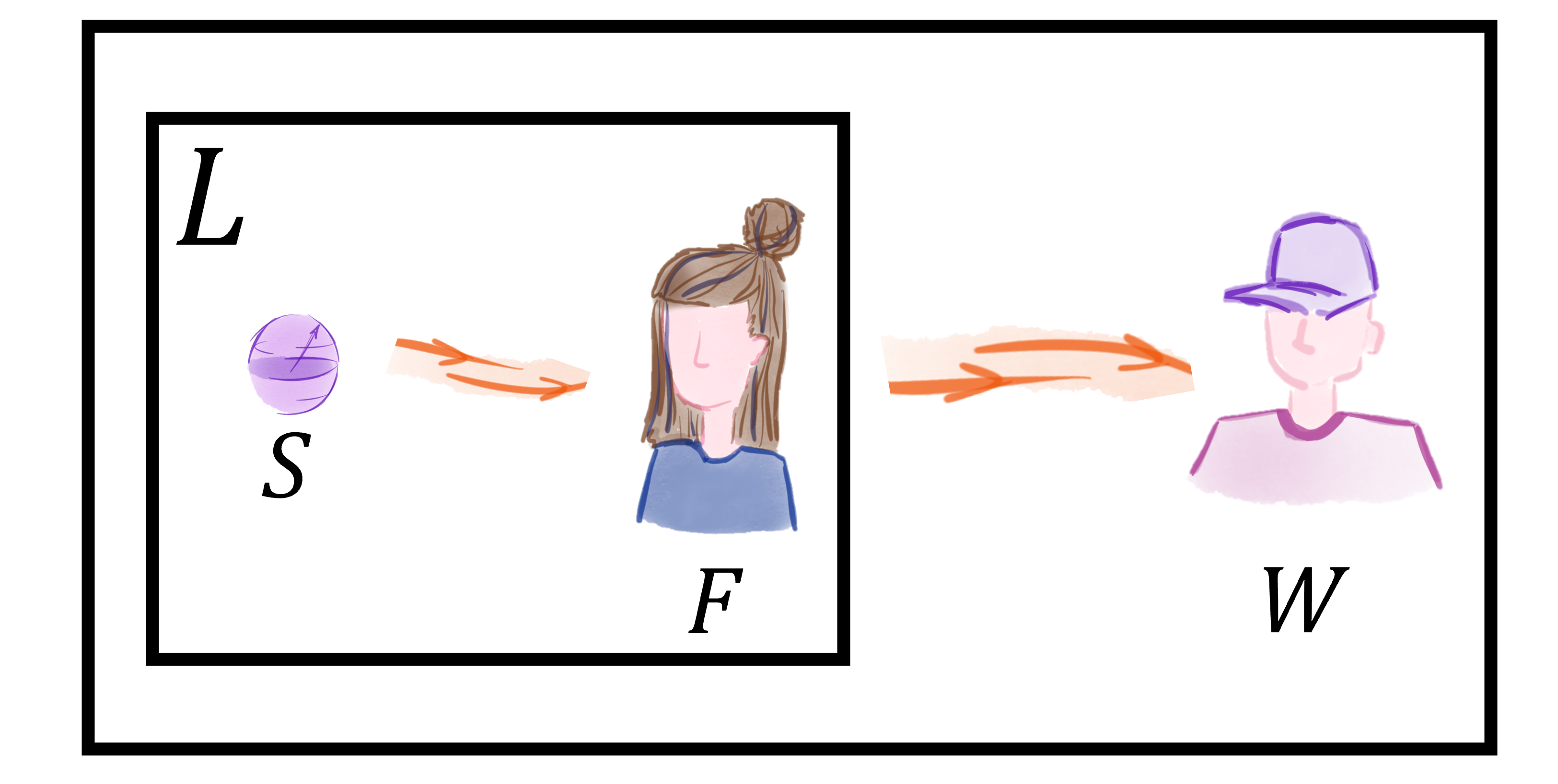}
    \includegraphics[width=0.45\linewidth]{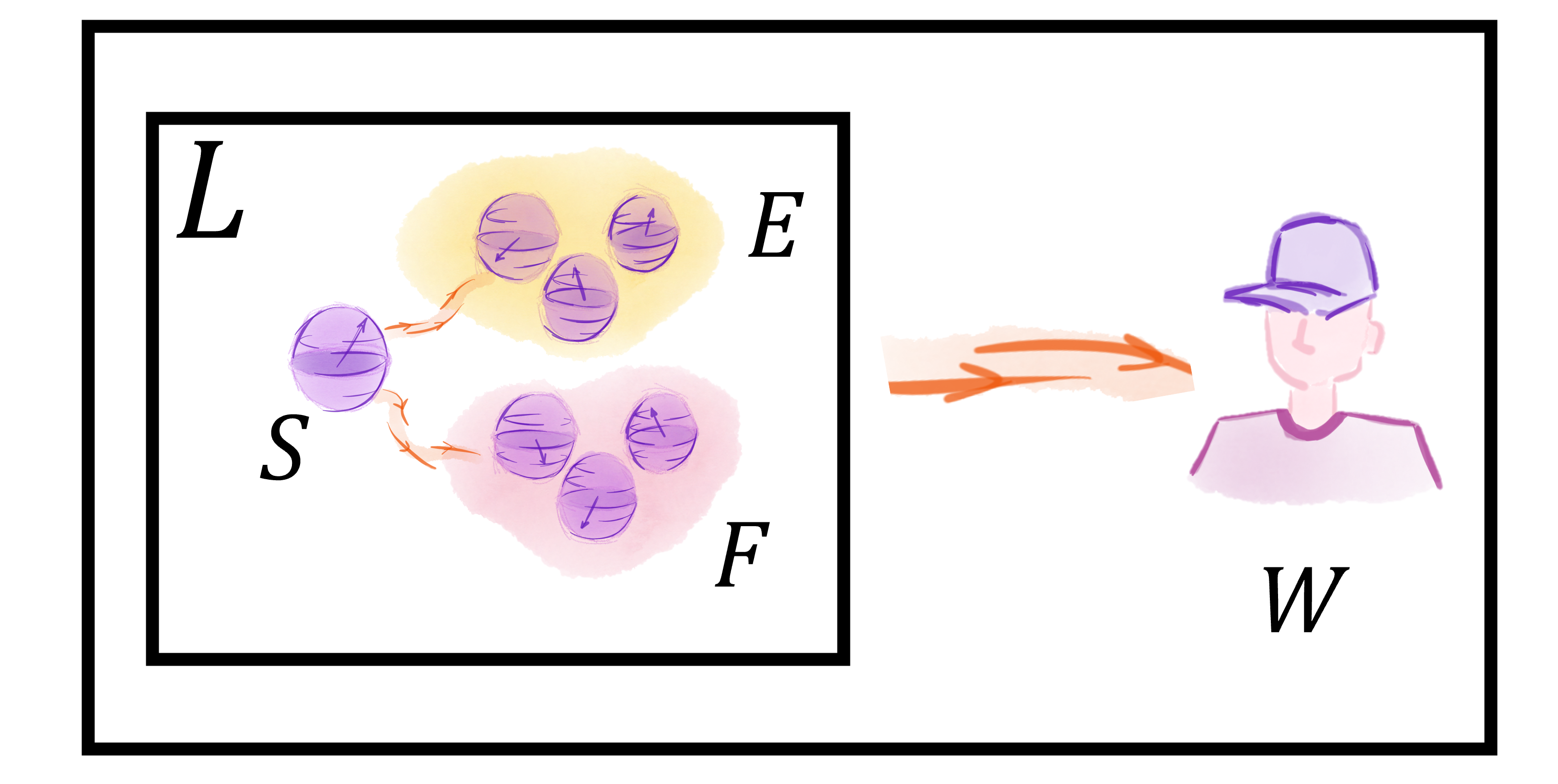}
 \caption{\textit{Left:} In the simple Wigner's Friend experiment of Sec.~\ref{ssec:The simple Wigner's Friend experiment1}, the Friend ($F$) is situated in an isolated Lab ($L$) and performs a measurement on a quantum system $S$. Wigner ($W$) is outside of this Lab and performs a measurement on the whole Lab. \textit{Right:} In the WF-QD setup of Sec.~\ref{ssec:WFandQD}, an environment is added and the Friend is modelled as a collection of qubits. }
\label{Fig:Wigner_simple}
\end{figure*}

\noindent The conventional Wigner's Friend (WF) thought experiment centres around a Friend, $F$, who sits in an isolated Lab, $L$, that also contains a quantum system, $S$ (the state of which is represented by the density matrix $\rho_S$). The Friend makes a measurement of the system $S$ corresponding to pointer basis, $\{\ket{i}\!\bra{i}_S\}_i$, of some observable (e.g. spin-up or spin-down for a $\sigma_z$ measurement). The key feature of a WF scenario is that the Friend is also represented by a quantum state $\rho_F$, meaning that the full quantum description of the Lab has a Hilbert space associated with each entity: $\mathcal{H}_L=\mathcal{H}_S\otimes\mathcal{H}_F$. We assume that it is possible to represent the quantum state of the Friend in a basis $\{\ket{f_i}\!\bra{f_i}_F\}_i$ on $\mathcal{H}_F$, where each projector $\ket{f_i}\!\bra{f_i}_F$ is associated with the Friend observing a particular measurement outcome $i$. In a conventional Wigner's Friend scenario, before the Friend's measurement the Lab state is usually assumed to be
\begin{equation}
    \label{eq:stateF_initial}
    \rho^{{\rm{init}}}_{L} =\rho_S\otimes\proj{r}_F =\sum_{ii'}\alpha_{ii'}\ketbra{i}{i'}_S\ox\proj{r}_F,
\end{equation}
where $\ket{r}$ is called the `ready-state' of the Friend, indicating that she has not performed her measurement yet.
Outside of the Lab sits Wigner, $W$, who is referred to as a `superobserver' (see Fig.~\ref{Fig:Wigner_simple}). From Wigner's perspective, the Friend's measurement is simply one part of the unitary evolution of the whole Lab. It is usually argued that the Friend, when seeing outcome $i$ in her measurement, will assign to the whole Lab the following `collapsed' state
\begin{equation}
    \label{eq:stateF_pure}
    \rho_L^F =\proj{i}_S\ox\proj{f_i}_F =\proj{i,f_i}_{SF}.
\end{equation}
By contrast, after the Friend's measurement, Wigner will assign to the Laboratory the state 
\begin{equation}
    \label{eq:stateW}
    \rho_L^W= U_F \rho_L^{{\rm{init}}}U_F^{\dagger}=\sum_{i,i'} \alpha_{ii'}\ket{i,f_i}\!\bra{i',f_{i'}}_{SF},
\end{equation}
where $U_F$ is the unitary evolution of L that happens when F makes her measurement. 

These different descriptions of a measurement by Wigner and his Friend and the resulting disagreeing state assignments are often referred to as the \emph{Wigner's Friend paradox}. However, up to this point there is no inconsistency, since both Wigner and the Friend would agree on the probabilities of the outcomes of the measurement performed. Formally, the Friend and Wigner assign the following probabilities to the possibility that the Friend's observed outcome $i$: 
\begin{equation}
\label{eq:P(f1)}
\begin{aligned}
P^F(i)&={\rm{Tr}}\left(\proj{i}_S\rho_S\right)=\lvert\alpha_{ii}\rvert^2\\
P^W(i)&= {\rm{Tr}}\left(\mathds{1}\ox\proj{f_i}_{F}\,\rho_{L}^W\right)=\lvert\alpha_{ii}\rvert^2,    
\end{aligned}
\end{equation}
where the projectors $\{\mathds{1}_S\ox\proj{f_i}_{F}\}_i$ are commonly referred to as Wigner's measurement in which he `asks his Friend what she observed'. (Throughout this work when discussing measurement outcomes, we will use superscripts to denote whose perspective the probabilities refer to, and indices such as $i$ to refer to the basis of the measurement.) The two expressions in Eq.~\eqref{eq:P(f1)} being the same means that the Friend's post-measurement state (encoding which outcomes she perceived) perfectly reflects the state of the system. If said expressions were not the same, it would imply that F is somehow not recording the measurement outcomes properly due to some measurement error.
We will refer to this effect as \textit{Broadcasting Error} for reasons that will become clear later.

The inconsistency arises when, after the Friend's measurement, instead of simply asking F what she observed, Wigner performs a `quantum measurement' on the whole Lab. (Note, this is often phrased in terms of Wigner `undoing' the Friend's measurement~\cite{schmid2023review}, but the two formulations are equivalent.) He is permitted to use arbitrary amounts of quantum control on the entire Lab Hilbert space, and thus can perform a measurement of the full state $\rho_{L}^W$, in any `pointer' basis $\{\ket{j}\!\bra{j}_{L}\}_j$.  Due to their different state assignments after the Friend's measurement, Wigner and his Friend will assign different probabilities to the outcomes $j$ of \textit{Wigner's} measurement:
\begin{equation}
\label{eq:P(w)}
\begin{aligned}
P^F(j)&= {\rm{Tr}}\left(\proj{j}_L\,\rho_L^F\right)\neq\\
P^W(j)&={\rm{Tr}}\left(\proj{j}_L\,\rho_L^W\right).
 \end{aligned}
\end{equation}
At most one of the probability assignments by Wigner and his Friend can agree with the actual frequencies one would observe in a real experiment. A WF setup can, hence, be used to decide whose state assignment -- $\rho_L^F$ or $\rho_L^W$ -- and, hence, whose description of the Friend's measurement was correct.\\

Wigner's measurement will in general alter the Lab state, and hence the observational state of the Friend~\cite{aaronson2018s}. It is further well-known that the measurement result the Friend initially observed is inaccessible to both herself and Wigner once the latter has performed his measurement~\cite{21AllardPhilippeBaumann,baumann2024wigner}. Hence, one might argue that the Friend should not condition on the specific outcome she observed in a WF scenario. As such it may make more sense for the Friend to assign a classical probabilistic mixture to the state of the Lab -- we say that the Friend knows a measurement has occurred but conditions only the probabilities of each outcome happening. In that case, when observing a definite outcome -- a fact that both she and Wigner can have access to throughout the experiment -- the Friend would assign the Lab the `decohered' state 
\begin{equation}
    \label{eq:stateF_mixed}
    \rho^{{\rm{Dec}}}_L=\sum_i |\alpha_{ii}|^2 \proj{i,f_i}_{SF},
\end{equation}
instead of the state $\rho_L^F$ in Eq.\eqref{eq:stateF_pure}. Note that $|\alpha_{ii}|^2=P^F(i)$ -- the Friend's prediction for the probability of obtaining outcome $i$ in her measurement. This, however, does not resolve the paradox. The disagreement between her and Wigner remains, since
\begin{equation}
      \label{eq:P^F(w)_mixed}
\begin{aligned}
 P^F(j)&=  {\rm{Tr}}\left(\proj{j}_L\,\rho^{{\rm{Dec}}}_L\right)\neq\\
 P^W(j)&={\rm{Tr}}\left(\proj{j}_L\,\rho_L^W\right) \, ,
\end{aligned}
\end{equation}
for a suitable measurement $\{\ket{j}\!\bra{j}_{L}\}_j$. The assigned Lab states after the Friend's measurement are still different for the two observers (decohered according to the Friend, containing coherence terms according to Wigner) as are their probability assignments for the outcomes of Wigner's measurement. This description of the conventional Wigner's Friend `paradox' is the first of many WF models we will describe in this work. We summarise all the different scenarios studied here in Tab.~\ref{tab:AllTheWFs}.\\

\begin{table*}
    \centering
    \renewcommand{\arraystretch}{1.4} % Increases row height for readability
    \begin{tabular}{|L{3cm}|L{1.75cm}|L{4cm}|L{3.5cm}|}
         \hline 
         \textbf{Name of model} & \textbf{Label} & \textbf{Description} & \textbf{Discussed in Section}\\\hline
         Simple Wigner's Friend scenario & WF & One superobserver, one observer, no environment & \ref{ssec:The simple Wigner's Friend experiment1} \\\hline
         Simple Wigner's Friend scenario with an environment & WF-E & One superobserver, one observer, an unspecified environment & \ref{ssec:WFwenv} \\\hline
         Simple Wigner's Friend scenario with quantum Darwinism & WF-QD & One superobserver, one observer, a specific model of an environment & \ref{ssec:WFandQD} \\\hline
         Extended Wigner's Friend scenario & EWFS & Two superobservers, two observers, no environment & \ref{ssec:LF} \\\hline
         Extended Wigner's Friend scenario with an environment & EWFS-E & Two superobservers, two observers, an unspecified environment & \ref{ssec:LF} \\\hline
         Extended Wigner's Friend scenario with quantum Darwinism & EWFS-QD & Two superobservers, two observers, a specific model of environments & \ref{ssec:Extended Wigner's Friend experiments2} \\\hline
    \end{tabular}
    \caption{A summary of all the different Wigner's Friend models discussed in this work.}
    \label{tab:AllTheWFs}
\end{table*}

\subsection{Wigner's Friend with an environment}
\label{ssec:WFwenv}

\noindent In the conventional Wigner's Friend setup outlined above, consider adding to the Lab an environment E in a Hilbert space $\mathcal{H}_E$, such that $\mathcal{H}_L=\mathcal{H}_S\otimes\mathcal{H}_F\otimes\mathcal{H}_E$. The purpose of this is to observe whether it `resolves' the disagreement between W and F. Some have argued that the Wigner's Friend effect is solely a result of the environment inside the Lab not being properly considered \cite{24ZukowskiMarkiewicz}. By proposing a specific model for how an environment could impact Wigner's ability to produce disagreements with the Friend, we can test to see whether this ability is solely dependent on the specific simplifications used in the usual model, or whether it persists even with the added complication of an environment.

Decoherence theory asserts that environment states will be commensurate with certain measurement outcomes being observed, similar to Friend states in the previous section. As such, according to Wigner's unitary description, when a decohering environment is also present, after the Friend's measurement the Lab state then becomes
\begin{equation}
    \label{eq:stateLab}
    \rho_{L}^W=\sum_{i,i'} \alpha_{ii'}\ket{i,f_i,e_i}\!\bra{i',f_{i'},e_{i'}}_{SFE},
\end{equation}
where the $\ket{e_i}$ are the environment states associated with each outcome $i$, that are assumed to be perfectly distinguishable once the measurement is completed. In~\cite{20Relano}, it is argued that that the reason for the Friend assigning the state in Eq.~\eqref{eq:stateF_mixed} is that the Friend does not have access to the environment, $\rho^{{\rm{Dec}}}_{L}=\rm{Tr}_E(\rho_L^W)$. (There it is implicitly assumed that $\rho_L^W=\rho_L$, i.e.\ that Wigner's assignment is the `correct' state.) This further implies that Wigner and his Friend will agree on the probabilities for any measurement by Wigner of only the system and the Friend:
\begin{align}
    \label{eq:deco_agreement}
     P^F(j)&= 
 {\rm{Tr}}\left(\proj{j}_{SF}\,\rho^{{\rm{Dec}}}_L\right) \\
&={\rm{Tr}}\left(\proj{j}_{SF}\ox \mathds{1}_E\,\rho_L^W\right)= P^W(j). \nonumber
\end{align}
Hence, assuming that Wigner (just like his Friend) \emph{cannot} have access to the environment `resolves' the Wigner's Friend paradox. In any realistic scenario, like a typical physics laboratory, an outside observer would not have coherent control over the unimaginably large Hilbert space for an environment. However, Wigner's Friend arguments are usually concerned with the \textit{in principle} problem of what would happen if Wigner somehow did have control over the entire Lab, including the environment, and so we continue under that assumption.

In general, provided that Wigner's measurement is on the whole Lab space $\mathcal{H}_L$, the Friend must still assign \textit{some} state to the environment when making predictions about that measurement. For the comparison between Wigner's prediction and the Friend's to make sense, they must be comparing like-for-like, meaning they must be looking at Hilbert spaces of the same size. Conditioning on the fact that she has performed a measurement, the Friend will assign the following Lab state, a replacement of Eq.~\eqref{eq:stateF_pure}:
\begin{align}
  \label{eq:stateF_LAb}
     \rho_{L}^F&=\sum_{i} |\alpha_{ii}|^2\proj{i}_S\ox\proj{f_i}_{F}\ox \rho^{\left(i\right)}_E,
 \end{align}   
where the $\rho^{(i)}_E$ represent the Friend's guess for an environment state. It will be based on her (limited) information about the the environment in her Lab. The optimal assignment, in the sense that it produces the correct diagonal elements of $\rho_L$ in the pointer basis, occurs for $\rho^{\left(i\right)}_E=\proj{e_i}_E$. However, an important fact that we assert here is that the Friend correctly guessing the environment states is not enough to resolve the paradox. This is because Wigner still assigns the state in Eq.~\eqref{eq:stateLab}, and for a general measurement we again obtain a disagreement like in Eq.~\eqref{eq:P^F(w)_mixed}.

In fact the disagreement between Wigner and the Friend is due to several effects. The first effect is the Friend's general lack of information about the environment, which means she is unable to predict the outcomes of measurements on the whole Lab like Wigner does. We refer to this effect as \textit{Classical Ignorance Errors}, since this disagreement between W and F would also arise in a purely classical version of the scenario (compare with~\cite{24JonesMueller}, a recent work that reproduces WF-like effects in purely classical scenarios). This effect is minimal if the Friend happens to assign the actual environment states $\ket{e_i}$.

The more interesting effect is F's lack of information about the \emph{coherences} in Wigner's state assignment. This effect is inherently quantum and will lead to a disagreement in the Friend's value of $P^F(j)$ even if she makes the optimal state assignment $\rho^{(i)}_E=\proj{e_i}$. We label this a \textit{True WF-Type Effect}.

A further possible effect is what we refer to as \textit{Control Error}. This corresponds to the case where Wigner is constrained in his ability to apply arbitrary unitaries or projectors on the Lab Hilbert space -- possibly due to lack of access to the environment, or to thermodynamic constraints (see~\cite{24DebarbaHuberFriis} for a study of how thermodynamic constraints can hinder a broadcasting process). This type of error limits Wigner's superobserver capabilities and will not be investigated further in this paper. 

Along with these effects and the Broadcasting Error mentioned earlier, we will also discuss one more confounding effect later -- see Tab.~\ref{tab:EffectsTable} for a full summary.

\subsection{Wigner's Friend in terms of quantum Darwinism}
\label{ssec:WFandQD}

\noindent The QD framework naturally accords itself with the Wigner's Friend scenario (other works have alluded to the connection before~\cite{22Zurek}), as both are concerned with the nature of information accessible to different observers, and the circumstances under which different observers agree on measurement outcomes. In our model, the decoherence process that leads to a QD-like state in the Lab between the system and the environment can be thought of as the Friend's initial measurement. Connecting the decoherence process in quantum Darwinism to the measurement process is a conceptual leap, but not one without precedent~\cite{18Zurek,23SchwarzhansBinderHuber,engineer2024equilibration}. The Friend is then considered to be just one fraction of the wider environment that contains information about the system, and Wigner the superobserver is treated as an external agent who can extract information from the environment inside the Lab (see Fig.~\ref{Fig:Wigner_simple}).

It has been shown \cite{19LeOlayaCastro} that a variant of QD called \textit{strong} quantum Darwinism, plus an additional assumption called strong independence, necessarily implies that the post-measurement state must have Spectrum Broadcast Structure:
\begin{equation}
    \label{eq:rhoSBS}
    \rho_{\rm{SBS}}=\sum_{i}p_i \ket{i}\!\bra{i}_S \bigotimes_{k=1}^N \rho_{k}^{\left(i\right)},
\end{equation}
where $p_i$ is the probability of outcome $i$ for a measurement of the system in basis $\ket{i}\!\bra{i}$ as predicted by quantum theory, and $\rho_k^{\left(i\right)}$ is the quantum state associated with a fraction of the environment labelled $k$ that is commensurate with having observed outcome $i$. For SBS states, the environment states associated with different measurement outcomes $i$ on a quantum system are orthogonal to each other:
\begin{equation}
    \label{eq:ortho}
    {\rm{Tr}}\left(\rho_F^{\left(i_1\right)}\rho_F^{\left(i_2\right)}\right)=0\ \ \forall i_1\neq i_2.
\end{equation}
This requirement ensures \emph{objectivity}~\cite{15HorodeckiKorbiczHorodecki},
the notion that multiple observers agree on the outcomes of the measurement. An SBS state is fully decohered and the information about the pointer basis $i$ of the system is perfectly encoded in every part of the environment. If we describe the Friend's measurement in terms of QD and require that the state of her Lab is given by Eq.~\eqref{eq:rhoSBS}, there is no Wigner's Friend paradox. Both Wigner and the Friend assign a perfectly decohered Lab state after the Friend's measurement, the only possible difference being the Friend assigning a different state to the environment due to a lack of access to the environment inside her Lab. Hence, any disagreement between Wigner and his Friend can be attributed to a Classical Ignorance Error. 

Here, when applying QD to Wigner's Friend scenarios, we assume that the measurement the Friend performs is a QD process that does not reach an SBS state exactly~\cite{23SchwarzhansBinderHuber}. We therefore consider \emph{SBS-like} states instead of the strict SBS state in Eq.~\eqref{eq:rhoSBS}, and use this to construct the QD analog of the Wigner's Friend scenario introduced in Sec.~\ref{ssec:The simple Wigner's Friend experiment1}. We assign some environment states $\rho_F^{\left(i\right)}$ to represent the Friend observing outcome $i$. The rest of the states then represent a residual environment also `observing' outcome $i$, $\rho_E^{\left(i\right)}$. Hence, the Friend's measurement results in the Lab being in an SBS-like state:
\begin{align}
    \label{eq:rhoLab1}
    \rho_{L}&=\sum_{i}q_i \ket{i}\!\bra{i}_S \bigotimes_{k=1}^N \rho_{k}^{\prime\left(i\right)} +\sigma\\
    &=\sum_{i}q_i \proj{i}_S \otimes \rho_F^{(i)}\otimes \rho_{E}^{(i)} +\sigma \, ,\nonumber
\end{align} 
where $\{\proj{i}_S\}_i$ is the pointer basis of the Friend's measurement, $q_i$ are the probabilities for measurement outcomes $i$ encoded in the post-measurement Lab state (not necessarily the same as the actual $p_i$ of the pre-measurement system) and $\rho_k^{\prime\left(i\right)}$ are environment states corresponding to different measurement outcomes $i$. 
As in~\cite{20Relano}, here we implicitly assume that the state of the Lab from Wigner's perspective is the correct one: $\rho_L=\rho_L^W$.
If this were an actual SBS state, then the overlap from Eq.~\eqref{eq:ortho} would be zero for all $i_1\neq i_2$ (and similar for $\rho_E^{\left(i\right)}$), and the overlap $\sigma$ between different classical outcomes would vanish.

We refer to any non-zero magnitude of the overlap in Eq.~\eqref{eq:ortho} as the \emph{non-objectivity} of the Friend. Classically, we expect that there should be no non-objectivity -- all observers should agree on measurement outcomes. Hence, in our model, the analog of the Friend assigning the decohered state to the Lab is the Friend assigning an exact SBS state to the Lab, where the non-objectivity terms are exactly zero:
\begin{equation}
\rho_L^F=\rho_{{\rm{SBS}}}.
\end{equation}
(Note that our use of the term `non-objectivity' here is non-standard. More formally, what we label the `non-objectivity' is merely one prerequisite for the actual presence of non-objectivity in the QD sense -- see for instance~\cite{23ChisholmInnocentiPalma} for a recent work discussing objectivity in QD.)

The above Eq.~\eqref{eq:rhoLab1} simplifies greatly if the equilibration process describing the Friend's measurement is governed by a broadcasting Hamiltonian:
\begin{align}
    H_L&=\sum_i \proj{i}\ox \sum_{k=1}^N c_k H^{(i)}_k \nonumber\\
    &= \sum_i \proj{i}\ox c_FH_{F}^{(i)}\ox c_EH_E^{(i)}.
    \label{eq:Hamiltonian_Friend}
\end{align}
Utilising ideas from the literature on equilibration on average~\cite{09LindenPopescuShort,11Short,23SchwarzhansBinderHuber}, we assume that this post-decoherence state of the Lab after the Friend's measurement can be calculated using a pinching map~\cite{20Romero,engineer2024equilibration}, which is known to be equivalent to a full dephasing map that deletes all coherences in the pointer basis (up to complications surrounding degeneracies which we ignore here). If the Hamiltonian $H_L$ has eigenvalues and eigenvectors $\{E_n,\Pi^H_n\}_{n}$, then the pinching map of the initial state $\rho_L^{{\rm{init}}}$ is given by
\begin{equation}
\rho_L=\sum_n\Pi_n^H\rho_L^{{\rm{init}}}\Pi_n^H,
\label{eq:PinchingMap}
\end{equation}
which is what we use for the numerical calculations in Secs.~\ref{ssec:Numerical results1} and~\ref{ssec:Numerical results2}.
This causes the $\sigma$ in Eq.~\eqref{eq:rhoLab1} to vanish. We also have that $q_i={\rm{Tr}}\left(\proj{i}_S\,\rho_S\right)=p_i$ (compare Eq.~\eqref{eq:rhoSBS} and see~\cite{23SchwarzhansBinderHuber,engineer2024equilibration}), meaning that the Broadcasting Errors mentioned earlier (see Tab.~\ref{tab:EffectsTable}) are not present. In this work we only focus on broadcasting Hamiltonians that do not permit Broadcasting Errors, and so can assume that the post-measurement state of the Lab is of the form
\begin{align}
    \label{eq:rhoLab2}
    \rho_{L}&=\sum_{i}p_i \proj{i}_S \otimes \rho_F^{(i)}\otimes \rho_{E}^{(i)} \, ,
\end{align} 
whilst noting that the broadcasting Hamiltonian still allows for some \emph{small} non-objectivity, meaning that $0<\Tr \left(\rho_F^{(i_1)}\rho_F^{(i_2)}\right) \ll 1$. 

Additionally, Wigner is assumed to have complete access to the Lab state in Eq.~\eqref{eq:rhoLab2}, while the Friend only knows her observed result and that her measurement has been performed. (If Wigner did not have complete access to the Lab state or was unable to perform arbitrary unitaries or projectors on it, this would correspond to Control Errors as discussed before and in Tab.~\ref{tab:EffectsTable}.) 

Next, the Friend assigns an SBS state to the Lab, which involves assigning some states to the environment components. If the Friend's choice for the states to assign the environment is sub-optimal, then she will suffer from Classical Ignorance Errors when making her predictions for W's measurements, which serves as an obfuscating, or confounding effect when trying to observe WF-type disagreements. As such, to minimise this source of error, we allow the Friend to make the best guess for a Lab SBS state that she can possibly make:
\begin{align}
    \label{eq:rhoLabF}
    \rho_L^F&=\sum_{i}p_i \proj{i}_S \otimes \frac{\Pi_F^{(i)}}{\Tr\left(\Pi_F^{(i)}\right)}\otimes \frac{\Pi_{E}^{(i)}}{\Tr\left(\Pi_{E}^{(i)}\right)} \, ,
\end{align} 
where the $\Pi_F$ and $\Pi_{E}$ represent the optimal effective description by the Friend of herself and her Lab's environment~\cite{08Montanaro,15BrandaoPianiHorodecki,mironowicz2017monitoring,22PoderiniRodariMoreno,engineer2024equilibration}. 
These operators most accurately reproduce the measurement statistics of system S when applied to the respective fraction of the environment and are therefore called \emph{`optimal projectors'} in~\cite{engineer2024equilibration} (based on a similar expression in \cite{15BrandaoPianiHorodecki, 22PoderiniRodariMoreno} -- the latter of which also briefly mentions possible connections to Extended Wigner's Friend Scenarios). Note, however, that in general $\Pi_F$ and $\Pi_{E}$ may be POVMs instead of projectors. We will return to these operators in much more detail later, as they also represent Wigner's best choice of measurement operator to observe the Friend's measurement outcomes.

Importantly, in our model, Wigner cannot perfectly reproduce the measurement statistics of the system by `observing' the quantum state of the Friend, since we have:
\begin{align}
    \label{eq:WF1}
   P^F(i)&=\Tr \left(\proj{i}\rho_S \right)\nonumber\\
   &= \Tr \left(\left(\proj{i}_S \ox \mathds{1}\right) \rho_L^F \right)\nonumber\\
   &=p_i,\nonumber\\
   P^{W}(i) &= \Tr \left( \left(\mathds{1}\otimes \Pi_F^{(i)}\right)\rho_L\right)\nonumber\\
   &= \sum_{k} p_k \Tr\left(\Pi_{F}^{i} \rho_{F}^{(k)}\right)\\
   &\neq P^F(i) . \nonumber
\end{align} 
In contrast to the conventional Wigner's Friend scenario where Wigner can simply `open the box' and ask the Friend what she observed, here we no longer have perfect agreement between Wigner and his Friend when they predict the Friend's observed result. (See~\cite{23XuSteinbergNguyen} for another recent work that  considered constraints on Wigner's ability to observe the Friend's measurement outcomes, and see~\cite{moreno2022events} for a similar discussion in the extended case.) We refer to this source of potential error as \textit{State Discrimination Errors} (see Tab.~\ref{tab:EffectsTable}), since the problem ultimately reduces to Wigner the superobserver being constrained by quantum state discrimination limitations when trying to distinguish $\rho_F^{\left(i_1\right)}$ from $\rho_F^{\left(i_2\right)}$ to obtain the correct value of $P^F(i)$, see~\cite{09BarnettCroke}.

This State Discrimination Error serves as a confounding effect in any attempt to see WF-type effects in a QD framework, and so when discussing Wigner's Friend setups we need to first set a limit for the disagreement in Eq.~\eqref{eq:WF1} we are willing to accept. Here we assign a parameter $\epsilon$: 
\begin{equation}
    \label{eq:epsilondef1}
    \epsilon\coloneq|P^{W}(i)-P^F(i)|,
\end{equation} 
and assert that witnessing true WF-type effects requires $\epsilon\ll 1$. If $\epsilon$ is too large, then we say that no definite result was obtained during the interaction between the quantum system and the Lab environment (including the Friend). 

Besides the State Discrimination Error, we are also interested in modelling in a QD context the Classical Ignorance Error mentioned earlier. As such, for comparison, we will also consider what we call the `Bad Friend' case, where the Friend has no effective description of the Lab environment. When assigning the Lab state, she thus assigns the environment the maximally-mixed state:
\begin{align}
    \label{eq:rhoLabBadFriend}
    \rho_L^B&=\sum_{i}p_i \proj{i}_S \otimes \frac{\Pi_F^{(i)}}{\Tr\left(\Pi_F^{(i)}\right)}\otimes \frac{1}{d_E} \mathds{1}_E \, .
\end{align} 
By comparing results for $\rho_L^B$ and $\rho_L^F$ we can see exactly how much of a difference the Classical Ignorance Error can make in a WF-QD scenario. 
A simple Wigner's Friend paradox now appears under the following circumstances. Let us consider a different measurement on the Lab, given by POVM elements $\{M_j\}_j$, for which Wigner and his Friend disagree in their probability assignments for outcome $j$:
\begin{align}
    \label{eq:WF2}
    P^{F}(j)&=\Tr \left(M_j\rho_L^F \right) 
    \neq \Tr \left(M_j\rho_{L} \right) =P^{W}(j)\, ,
\end{align} 
which will happen for some measurements by Wigner since the two states $\rho_L$ and $\rho_L^F$ assigned by Wigner and his Friend respectively are different. However, we have already accepted that disagreements between $P^F$ and $P^W$ up to $\epsilon$ are possible due to measurement inaccuracy, based on Eq.~\eqref{eq:WF1}. We therefore require that
\begin{align}
    \label{eq:QDparadox}
    \Delta \coloneq |P^{W}(j)-P^{F}(j)| \gg \epsilon
\end{align}
in order to consider the disagreement in Eq.~\eqref{eq:WF2} to be a Wigner's Friend paradox. As discussed in the following, we can indeed find such measurements numerically for the above measurement model.

\subsection{Implementing the WF-QD model}
\label{ssec:Numerical results1}

\noindent In order to test whether the specific QD model we propose allows for observing WF-type effects, and to investigate the potential emergence of classicality as the Friend and environment increase in size, we simulate the model numerically for up to 11 qubits. We refer to Appendix~\ref{ssec:TheMEHBit} for more details of how the numerics were implemented, and here mainly focus on the outcomes of the simulations.

\begin{figure*}[htbp]
\includegraphics[width=0.75\linewidth]{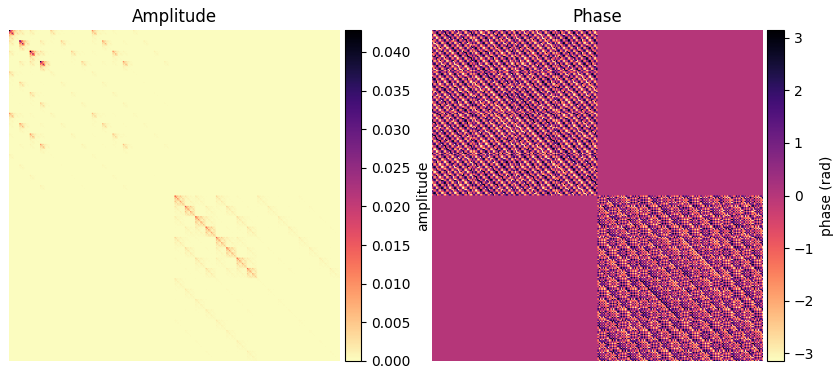}
 \caption{Density plot of an SBS-like state of the whole Laboratory, $\rho_L$, when the Friend's measurement is described by a decohering equilibration process governed by the broadcasting Hamiltonian in Eq.~\eqref{eq:Hamiltonian_Friend}. The two blocks correspond to the two pointer states $\ket{0}$ and $\ket{1}$ of the system. Here, $N_F$ and $N_E$ (the number of qubits that F and E are respectively composed of) are 3 and 4.}
\label{Fig:example_density}
\end{figure*}

Fig.~\ref{Fig:example_density} shows an example of a $\rho_L$ produced by our simulation, where $\rho_L$ is the post-measurement state of the Lab, containing the system, the Friend, and the environment. It has a distinctive block-diagonal structure: if the measurement on the system is done in the $\{\ket{0},\ket{1}\}$ computational basis, each on-diagonal block represents the system being in either the 0 or the 1 state, and the Friend and environment qubits also being in states commensurate with having observed the outcomes 0 or 1. In the simulations we performed, we sampled the Hamiltonian parameters randomly from the Gaussian Unitary Ensemble and averaged over the resulting outcomes (See Appendix~\ref{ssec:TheMEHBit}). Any given GUE sampling produces different-looking density plots of $\rho_L$, but the block-diagonal structure remains the same, as it is indicative of the broadcasting Hamiltonian used to produce it. 

Given this structure for $\rho_L$, we first ask whether Wigner is able to `ask the Friend what she saw', as is conventional in any WF scenario. In quantum Darwinism terms, this amounts to a state discrimination problem, and as mentioned before an inability to do this produces what we label State Discrimination Errors. The state $\rho_F^{\left(i\right)}$ is a collection of qubits containing information about measurement outcome $i$, and any observer trying to discern information about what state the system is in must attempt to distinguish this state from all the other $\rho_F^{\left(i'\right)}$. If $\rho_L$ was in a perfect SBS state, then all the $\rho_F^{\left(i\right)}$ would be perfectly distinguishable from each other and there would be no State Discrimination Errors, but the non-objectivity means that there is a chance that Wigner can be led to believe the Friend observed one outcome when in reality she observed a different one. 
Here we ask what operator Wigner can use to have the best chance of correctly guessing the right state, as in many previous investigations into quantum state discrimination~\cite{08Montanaro,mironowicz2017monitoring,18LeOlayaCastro,22PoderiniRodariMoreno,engineer2024equilibration}. In order to find the optimal operator to discriminate between more than two measurement outcomes, a numerical convex optimisation is required over all possible POVMs. In this work, however, we only consider situations where there are two possible measurement outcomes, which we label $0$ and $1$. In such cases the well-known Helstrom POVM~\cite{09BarnettCroke} can be used to optimally discern between the outcomes (as was done in \cite{mironowicz2017monitoring}).

Let ${\rm{Tr}}\left(\Pi_F^0\rho_F^0\right)=e_0$ and ${\rm{Tr}}\left(\Pi_F^1\rho_F^1\right)=e_1$, where $\Pi_F^{i}$ is the best possible projector Wigner can use to discern between the two possible Friend states. Also let $p_0$ and $p_1$ be the probabilities for the two outcomes of measurements in the computational basis as calculated for the system state. Then for Wigner's prediction for one of the Friend's two outcomes ($i=0$, say), we have 
\begin{equation}
    \begin{aligned}
    \label{eq:PWi0}
    P^W(i=0)&= {\rm{Tr}}\left(\left(\mathbb{1}_S\otimes \Pi_F^0\otimes\mathbb{1}_E\right)\rho_L\right)\\ &= p_0{\rm{Tr}}\left( \Pi_F^0\rho_F^0\right) + p_1{\rm{Tr}}\left(\Pi_F^0\rho_F^1\right).\\
    &=p_0e_0 + p_1{\rm{Tr}}\left(\left(\mathbb{1}_F-\Pi_F^1\right)\rho_F^1\right)\\
    &=p_0e_0 + p_1{\rm{Tr}}\left(\rho_F^1\right) -p_1e_1\\ 
    &=p_0e_0 + p_1 - p_1e_1.
    \end{aligned}
\end{equation}

If $p_0=p_1=0.5$, this reduces down to 
\begin{equation}
    \label{eq:PWi0part2}
    P^W(i=0)=p_0\left(1+e_0 - e_1\right),
\end{equation}
meaning that the $\epsilon$ parameter introduced in Eq.~\eqref{eq:epsilondef1} is merely a function of $e_0-e_1$.

\begin{figure*}[ht]
    \centering
    \includegraphics[width=0.45\linewidth]{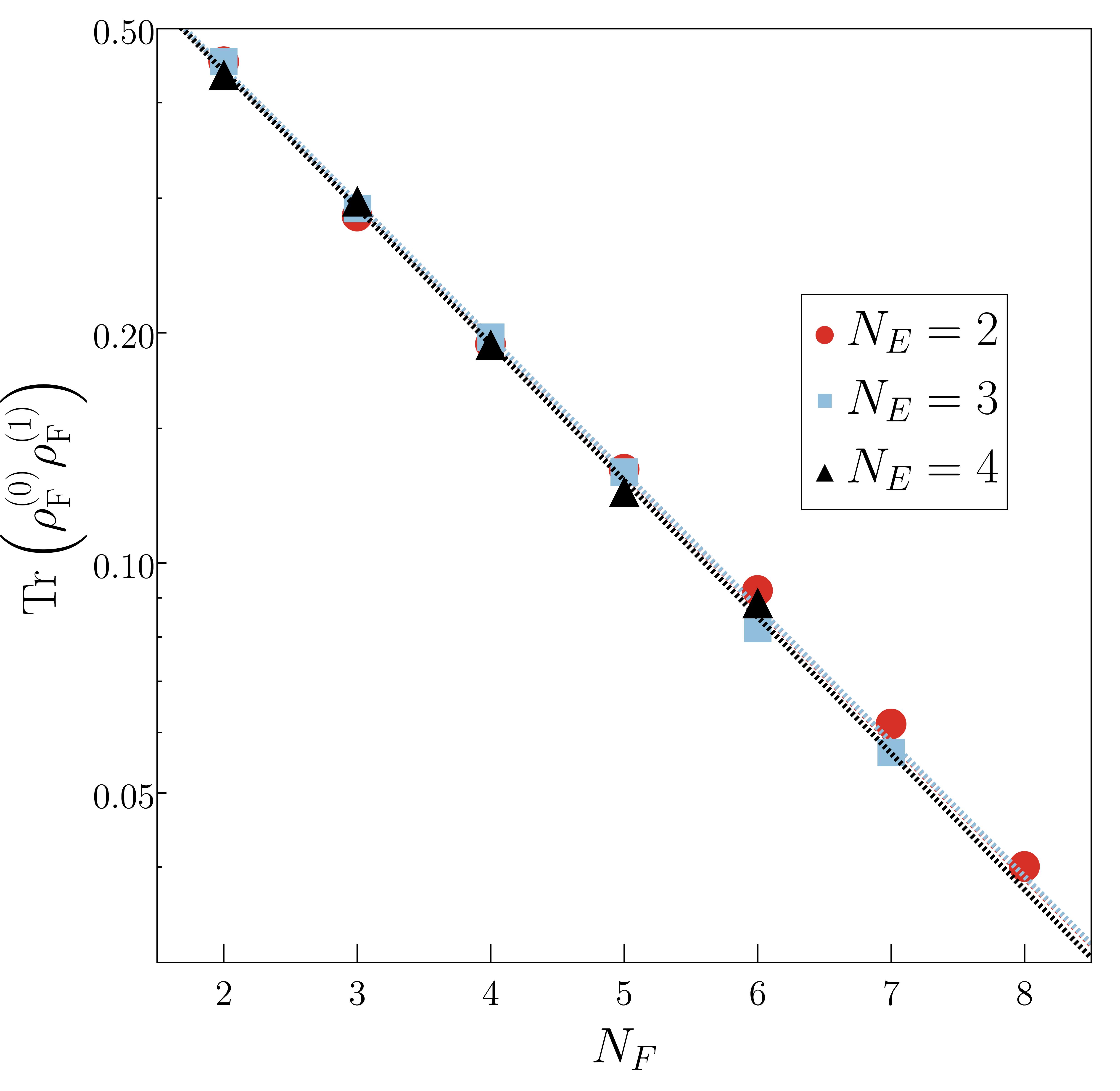}
    \includegraphics[width=0.45\linewidth]{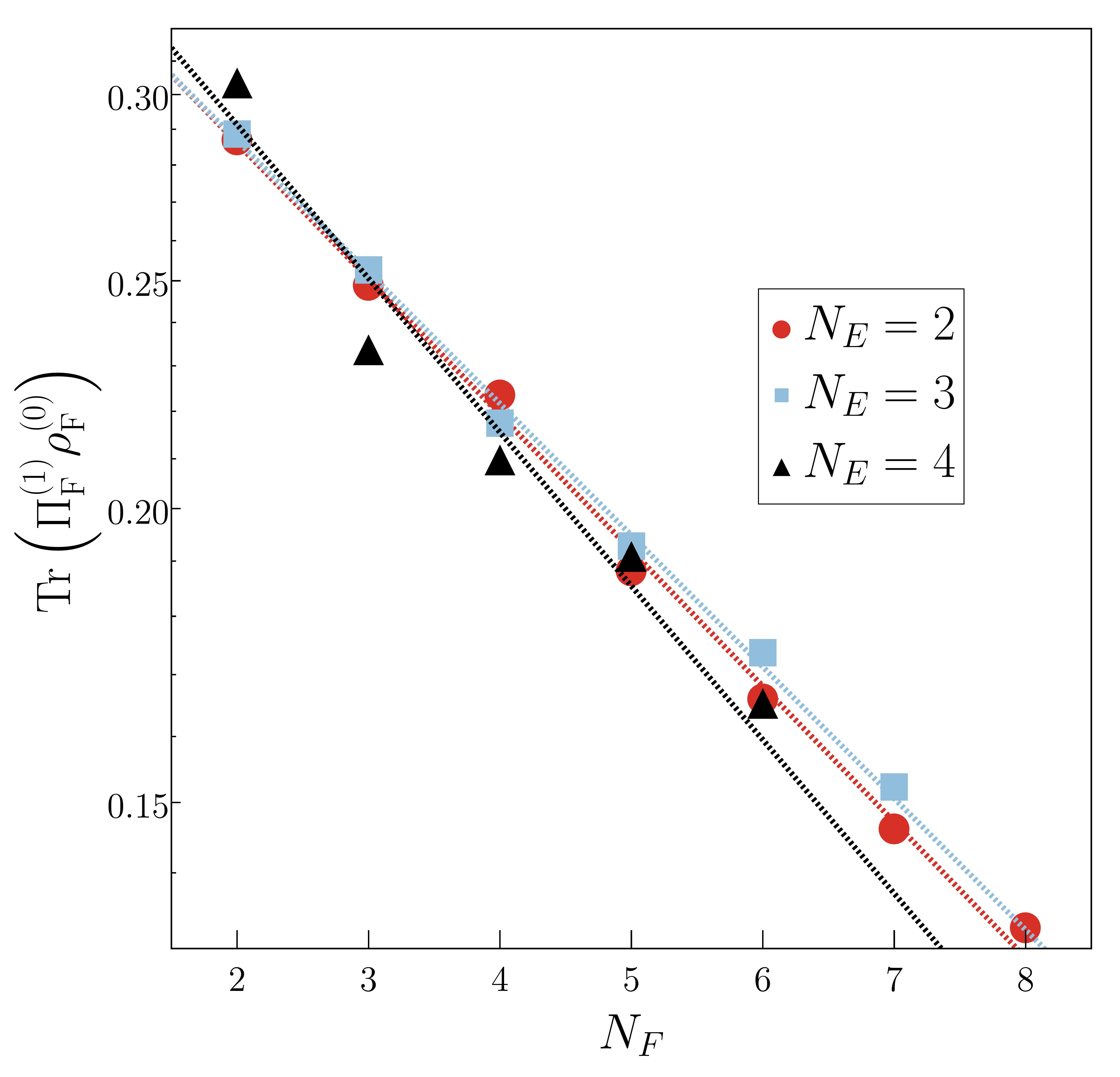}
    \caption{\textit{Left:} Plot (on a log scale) of a metric tracking the non-objectivity in the Friend, ${\rm{Tr}}\left(\rho_F^{0}\rho_F^{1}\right)$, as a function of $N_F$, for different values of $N_E$, when $p_0=p_1=0.5$. 200 GUE samples were used to generate the points in these plots (the SEM is negligible and hence error bars are not plotted). The logarithmic best-fit line is intended solely to guide the eye. \textit{Right:} A similar plot for the same scenario, but for ${\rm{Tr}}\left(\Pi_F^1\rho_F^{0}\right)$ (equivalent to $1-e_0$ in Eq.~\eqref{eq:PWi0} above). }
    \label{fig:OverlapsDecreasing}
\end{figure*}

Our simulations show that the non-objectivity is a size effect. The Friend and the environment within the Lab are represented by $N_F$ and $N_E$ qubits respectively, and in Fig.~\ref{fig:OverlapsDecreasing}, we see how the overlap between $\rho_F^{\left(0\right)}$ and $\rho_F^{\left(1\right)}$ decreases as $N_F$ increases (a similar result is seen but not shown here for $\rho_E^{\left(0\right)}$ and $\rho_E^{\left(1\right)}$ as $N_E$ increases). As a consequence, the quality of the optimal observables increase too. The overlap ${\rm{Tr}}\left(\Pi_F^1\rho_F^{0}\right)$ also decreases in size as $N_F$ increases -- meaning that Wigner becomes less likely to incorrectly guess the Friend's observed outcome.

Note that, so far, our results are merely reaffirmations of conventional quantum Darwinism -- that increasing Hilbert space dimension reduces non-objectivity. One can compare Fig.~\ref{fig:OverlapsDecreasing} to similar figures in~\cite{24StrasbergReinhardSchindler}, for instance.

For all data points generated in this work, the specific value was found to vary by a fairly large amount for any single sampling from the Gaussian Unitary Ensemble (see Appendix~\ref{ssec:TheMEHBit} for details). This led to notable standard deviations $\sigma_{{\rm{SD}}}$ for some of the points (particularly when $N_F$ and $N_E$ were small). But for all data points shown in the main text, generated using $N_{{\rm{GUE}}}=200$ GUE samplings, the standard error of the mean (SEM, calculated as $\sigma_{{\rm{SD}}}/\sqrt{N_{{\rm{GUE}}}}$) was found to be only marginally larger than the points themselves on the plots. This indicates the robustness of the results presented here.

\begin{figure*}[htbp]
    \centering
    \includegraphics[width=0.45\linewidth]{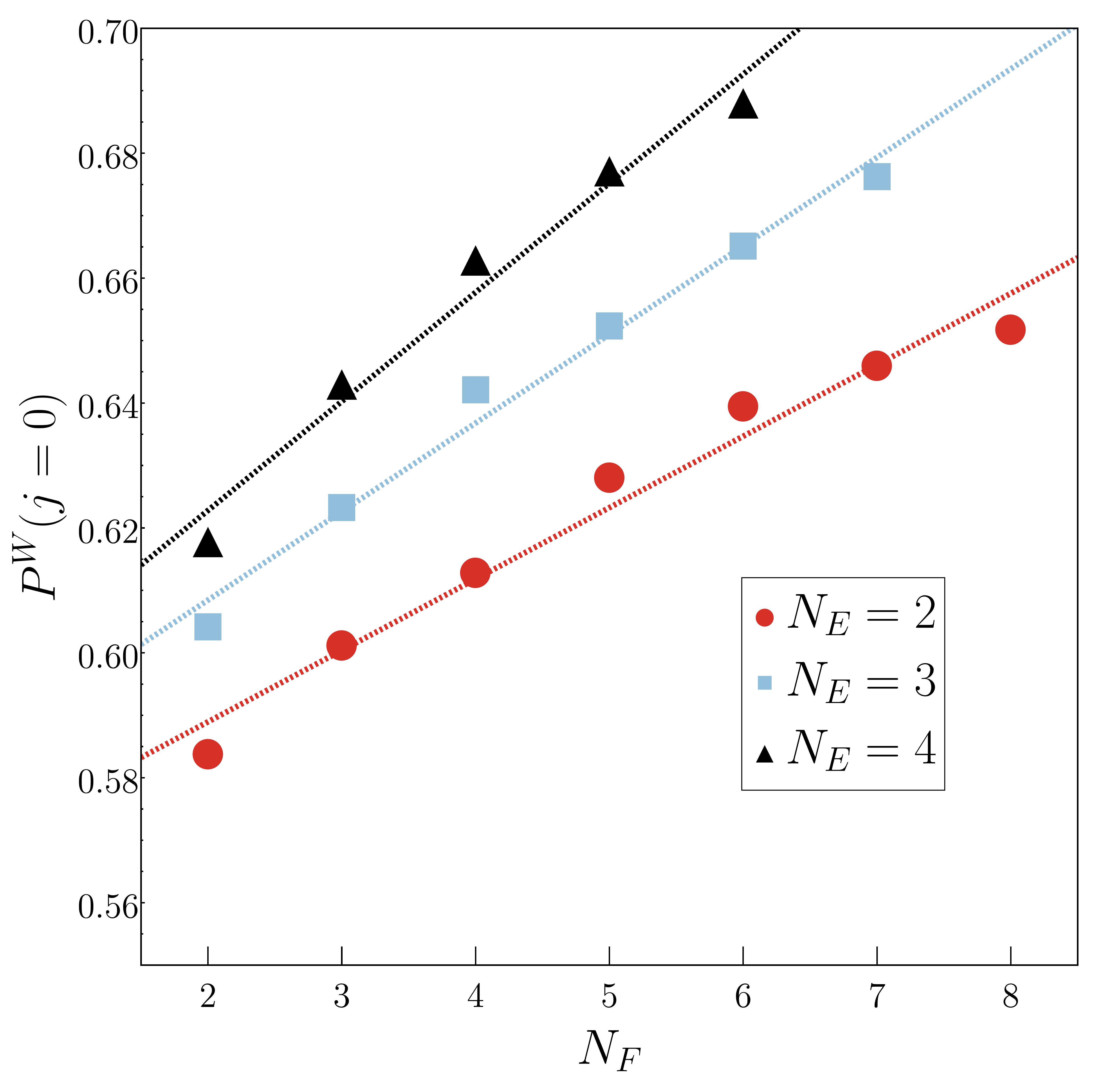}
    \includegraphics[width=0.45\linewidth]{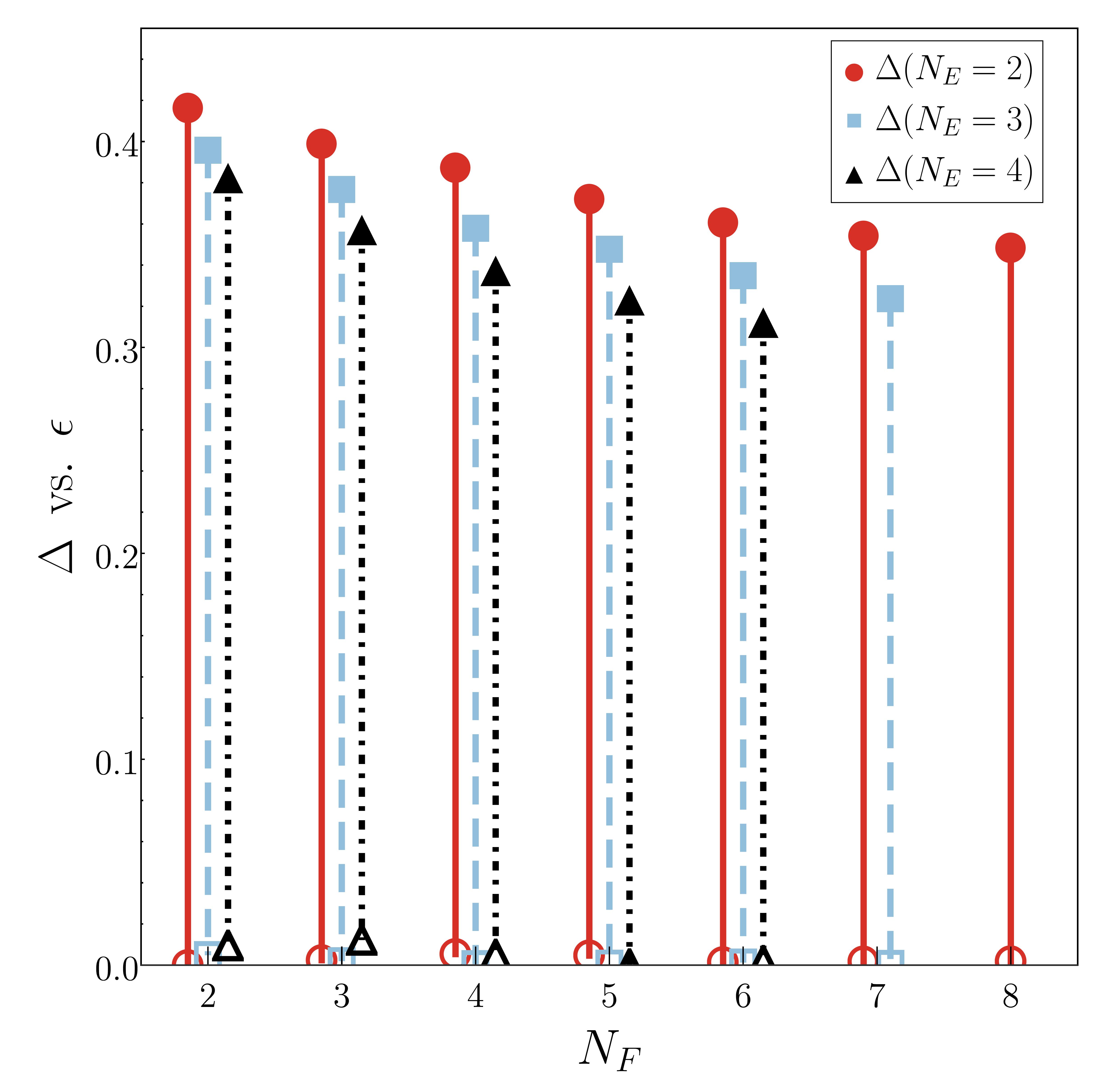}
    \caption{\textit{Left:} Plot of Wigner's measurement outcome $P^W(j=0)$ when using the POVM with elements $\{M_L^0,M_L^1\}$ (see Appendix~\ref{ssec:WignersPOVMs}), against $N_F$, for different values of $N_E$, when $p_0=p_1=0.5$. Plot generated using 200 GUE samples with negligible SEM. The best-fit lines are solely to guide the eye. \textit{Right:} Plot of the difference $\Delta=|P^W(j=0)-P^F(j=0)|$ (solid data points), compared to the difference $\epsilon=|P^W(i=0)-P^F(i=0)|$ (hollow data points). $\Delta >> \epsilon$ for all the $N_F$ and $N_E$ considered here signifies a genuine Wigner's Friend effect. Both larger $N_F$ and larger $N_E$ decrease the disagreement between W and F, hinting at an emergence of classicality.}
    \label{Fig:Plotepsilondelta}
\end{figure*}

After analysing the non-objectivity, we attempt to find a Wigner's Friend effect by generating large disagreement between $P^W(j)$ and $P^F(j)$ (large $\Delta$ -- see Eq.~\eqref{eq:QDparadox}). To do so, we choose a specific measurement basis $j$ for Wigner. We define a POVM composed of two elements $\{M_L^0,M_L^1\}$ acting on $\mathcal{H}_L$, which essentially asks whether the $F$ macrofraction records the same measurement outcome as the $E$ macrofraction (see Appendix~\ref{ssec:WignersPOVMs} for details). 
For this measurement, we find that the ``Bad Friend'' of Eq.~\eqref{eq:rhoLabBadFriend} consistently predicts drastically different results than the regular Friend. For $p_0=p_1=0.5$, we find that $P^B(j=0)=P^B(j=1)=0.5$, and the regular Friend using her best possible guess for the environment states consistently predicts $P^F(j=0)=1$ and $P^F(j=1)=0$ -- all with negligible standard deviation. Fig.~\ref{Fig:Plotepsilondelta} shows Wigner's observation of $j=0$, $P^W(j=0)$, as a function of $N_F$, for different values of $N_E$. It also shows the difference between $\epsilon$ and $\Delta$ defined in Eqs.~\eqref{eq:epsilondef1} and~\eqref{eq:QDparadox}.

The first thing we observe is that Wigner's value for $P^W(j=0)$ clearly differs significantly from the Friend's $P^F(j=0)$ (they lie well outside each others' $\sigma_{{\rm{SD}}}$). Next, we note that the disagreement between Wigner and his Friend for the measurement $M^j_L$ is much larger that that for the measurement $\Pi^i_F$ where Wigner `asks the Friend what she observed'. One can compare $\Delta$,
which varies from approximately 0.3 to 0.4, to $\epsilon$ varying between $10^{-5}$ and 0.012 (with negligibly small SEM). This is a strong indication that the goal has been satisfied of observing Wigner's Friend effects in the model presented in this work. However, we also observe a significant Classical Ignorance Error, as we find that the Bad Friend's predictions are as different as possible from the regular Friend's (with Wigner's predicted probabilities somewhere in between).

Moreover, there are hints towards the expected size effect occurring here -- our results indicate that, for larger $N_F$, it may be the case that Wigner's $P^W(j=0)$ gets closer to the Friend's $P^F(j=0)=1$. If this is the case, it implies that the WF-type effects disappear in the thermodynamic limit -- that they are a consequence of the small sizes of the systems involved. This could possibly indicate the emergence of classicality, and may have implications for research proposals that attempt to create WF-like scenarios involving larger quantum objects~\cite{23WisemanCavalcantiRieffel}. (These results should be compared with~\cite{zeng2024towards}, which implements Wigner's Friend scenarios on a quantum circuit with a similar number of qubits.) 

Recall also that the Bad Friend's values $P^B(j)$ do not vary with $N_F$ and $N_E$ -- as long as there is some environment the Friend lacks access to, she will make false predictions. This justifies our description of this as \textit{Classical} Ignorance -- the effect would be present in large and small systems. Hence, any future Wigner's Friend investigation on larger systems must also take these effects into account.

If $p_0\neq p_1$, the situation becomes less clear-cut. Much of the discussion of this situation we leave to Appendix~\ref{sec:puneven}. But we want to mention here that the uneven probability can make the Helstrom measurement's probability of success much lower~\cite{09BarnettCroke}, and hence $\epsilon$ much larger, as it becomes more prudent to simply guess you are in the higher-probability state most of the time.

\begin{table*}[ht]
    \centering
    \renewcommand{\arraystretch}{1.4} % Increases row height for readability
    \begin{tabular}{|L{2cm}|L{3cm}|L{6cm}|L{3cm}|}
        \hline
        \textbf{Effect} & \textbf{Variable it affects} & \textbf{Notes} & \textbf{In this work}\\
        \hline
        Broadcasting Errors & $P^F(i)$ & The Friend is somehow impeded in her attempt to make the initial measurement, and so her quantum state does not record the outcomes properly & Minimally present \\
        \hline
        State Discrimination Errors & $P^W(i)$ & Wigner is unable to read the Friend's quantum state perfectly to learn about her measurement outcomes & A notable but small source of error \\
        \hline
        Classical Ignorance Errors & $P^F(j)$ & The Friend makes a poor prediction for Wigner's measurement outcomes due to lack of access to the wider environment. (Compare values of $P^F(j)$ and $P^B(j)$) & A potentially large confounding factor\\
        \hline
        Control Errors & $P^W(j)$ & A general name for anything preventing Wigner from implementing his $j$-basis measurement or having arbitrary coherent control on $\mathcal{H}_L$ & Not modelled \\
        \hline
        \textbf{True WF-Type Effects} & $P^F(j)$ & \textbf{Wigner exploits coherences in} $\rho_L$ \textbf{that the Friend lacks access to, to make different predictions for his measurement outcome than the Friend} & \textbf{Detected in some cases}\\
        \hline
    \end{tabular}
    \caption{Summary of the different effects that could impact the measurement outcomes of Wigner and the Friend}
    \label{tab:EffectsTable}
\end{table*}

\section{Extended Wigner's Friend scenarios}
\label{sec:Extended Wigner's Friend scenarios1}

\subsection{The local Friendliness no-go theorem}
\label{ssec:LF}

\begin{figure*}[htbp]
    \centering
    \includegraphics[width=0.6\linewidth]{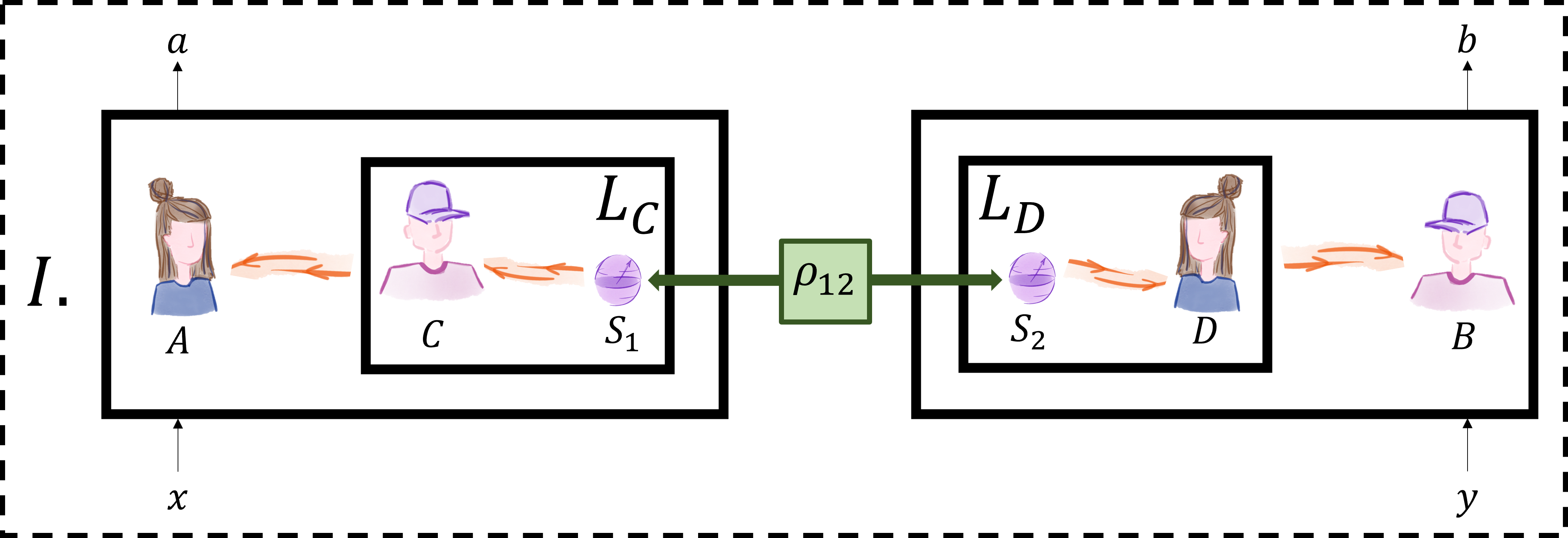} \\
    \vspace{0.1em}
    \includegraphics[width=0.6\linewidth]{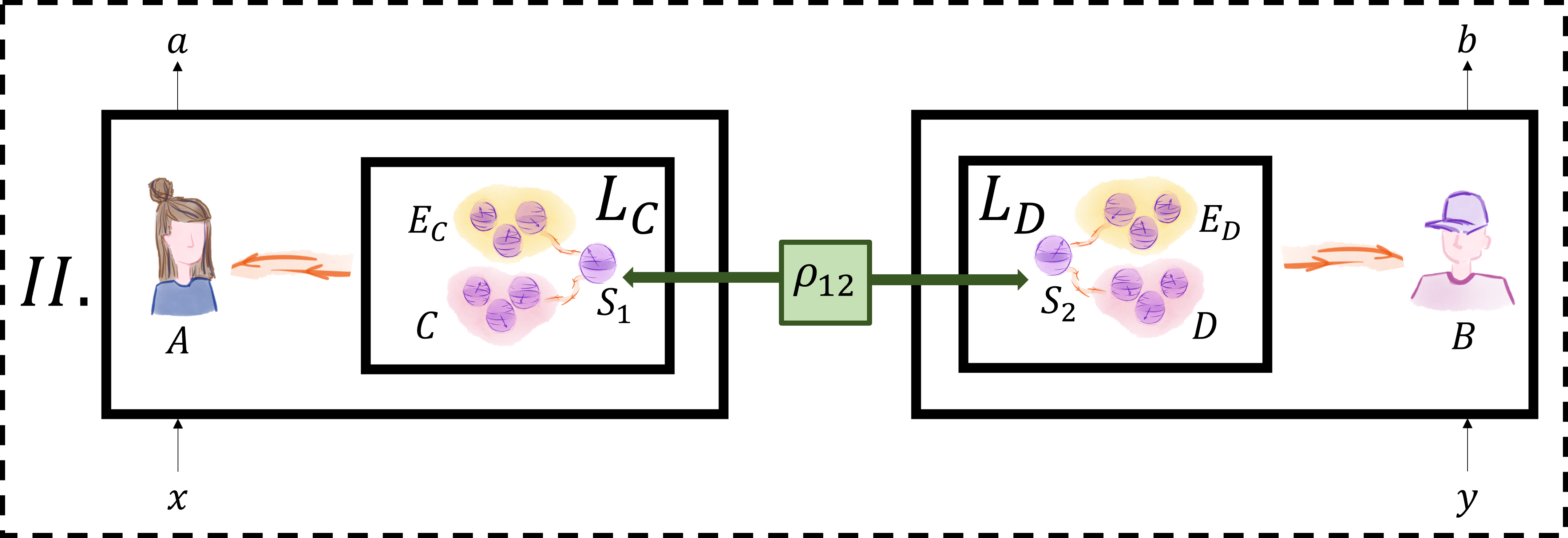}
    \caption{Extended Wigner's Friend experiments: Subfigure $I.$: As in Sec.~\ref{ssec:LF}, a bipartite quantum state is distributed between two Wigner's Friend setup. The Friends, Charlie ($C$) and Debbie ($D$), perform fixed measurements on their respective subsystem obtaining measurements results $c$ and $d$. The Wigners, Alice ($A$) and Bob ($B$), each choose among possible measurements according to their inputs $x$ and $y$ and observe results $a$ and $b$ respectively. Subfigure $II.$: The EWFS-QD setup from Sec.~\ref{ssec:Extended Wigner's Friend experiments2} with environments added and the Friends replaced with collections of qubits.}
    \label{Fig:Wigner_ext}
\end{figure*}

\noindent In a typical EWFS, two Friends, Charlie and Debbie, are each handed one half $\rho_1$ and $\rho_2$ of an entangled system $\rho_{12}$. They each measure their respective subsystem and are themselves measured by two superobservers, Alice and Bob respectively, who can choose between multiple measurement settings. In particular, in the setup depicted in Fig.~\ref{Fig:Wigner_ext}, $a$, $b$, $c$, $d$, $x$, and $y$ are possible values of random variables from distributions $A$, $B$, $C$, $D$, $X$, and $Y$ respectively. The random variable $X$ is an input accessible only to Alice, and it determines which measurement she performs (likewise for the variable $Y$ and Bob). For one of the inputs, say for $X=1$, Alice performs on Charlie's Lab the equivalent of the $i$-basis measurement in the previous section (`asking Charlie what he observed'), while for other values of $X$ she performs the equivalent of $j$-basis measurements (and equivalent for Bob and $Y$ acting on Debbie's Lab). 

The variable $A$ with possible values $a$ represents the outcomes of the measurements Alice performs on the Lab containing Charlie and his half of the system (and likewise for $B$ and Bob's measurement of Debbie's Lab with outcomes $b$). Furthermore, $C$ represents the outcomes of Charlie's measurement inside his Lab with possible values $c$ (likewise for $D$, $d$ and Debbie). As stated in~\cite{moreno2022events}, the local Friendliness assumptions for such setups are
\begin{itemize}
    \item{\textbf{Absoluteness of observed events (AOE)}\\
    There exists a probability distribution $P(a,b,c,d|x,y)$ such that:
    \begin{enumerate}
        \item{$P(a,b|x,y)=\sum_{c,d}P(a,b,c,d|x,y)$}
        \item{$P(a|c,d,x=1,y)=\delta_{ac}$ for all $a,c,y$
        $\Leftrightarrow P(a=c|d,x=1,y)=1$}
        \item{$P(b|c,d,x,y=1)=\delta_{bd}$ for all $b,d,x$
        $\Leftrightarrow P(b=d|c,x,y=1)=1$}
        
    \end{enumerate}}
    \item{\textbf{Local agency (LA)}\\
    Freely chosen settings are uncorrelated with anything outside of their future light cone:
    \begin{enumerate}
        \item{$P(c,d|x,y)=P(c,d)$ \newline (no super-determinism)}
        \item{$P(a|c,d,x,y)=P(a|c,d,x)$ \newline (no-signaling)}
        \item{$P(b|c,d,x,y)=P(b|c,d,y)$ \newline (no-signaling)}
    \end{enumerate}
    }
\end{itemize}
In contrast to the simple Wigner's Friend setup, we are now not concerned with how the two `Friends' would model the post measurement state but rather whether measurements \{$A_x$\} and \{$B_y$\} by Alice and Bob can violate so-called Local Friendliness inequalities. They are derived from the LF assumptions analogous to Bell Inequalities being derived from the assumptions of `local realism' -- the idea that every random variable has its value pre-determined before it is observed, and no signal can travel arbitrarily fast between observers.
Also analogous to Bell inequalities and local realism, when LF inequalities are violated the LF assumptions cannot be simultaneously be satisfied. This has been taken to mean that in EWFS there is no absolute notion of events and that the outcome of a measurement is only defined relative to an observer~\cite{di2021stable}. Note that it was shown in~\cite{20BongUtrerasAlaconGhafari} that in general Bell-local correlations are a strict subset of LF correlations and that there are correlations that violate Bell inequalities but still satisfy Local Friendliness. This results from the fact that the  LF assumptions above are strictly weaker than those made for Bell inequalities.

In the simplest case, Alice and Bob each choose from two possible two-outcome measurements. For this setup there exists an LF inequality  for the observed statistics of the two superobservers analogous to the CHSH inequality, for example,
\begin{align}
    \label{eq:LF-CHSH}
    &\langle{\rm{CHSH}}\rangle =\\ 
    &\langle A_0 B_0\rangle + \langle A_0 B_1\rangle - \langle A_1 B_0 \rangle + \langle A_1 B_1 \rangle  \leq 2, \nonumber
\end{align}
where $\langle A_x B_y \rangle =\sum_{a,b\in\{0,1\}}(-1)^{a+b} P(a,b|x,y)$. If the LF assumptions are satisfied, $\langle{\rm{CHSH}}\rangle$ is constrained to be less than 2. 

We now discuss the EWFS in Fig.~\ref{Fig:Wigner_ext} analogously to the simple setup in Sec.~\ref{ssec:The simple Wigner's Friend experiment1}. Let the source emit the state
\begin{align}
\label{eq:rho12}
    \rho_{12}= \sum_{kl} \gamma_{kl} \rho_1^{(k)}\ox \rho_2^{(l)},
\end{align}
for example, $\rho_{12}=\proj{\psi_{12}}$ with 
\begin{align}
\label{eq:psi12}
&\ket{\psi_{12}}=\\
&\frac{1}{\sqrt{2}}\left(\cos{\theta}(\ket{01}-\ket{10})-\sin{\theta}(\ket{00}+\ket{11})\right) \nonumber
\end{align}
as in~\cite{brukner2018no}. If the measurements by Charlie and Debbie are described unitarily, compare Eq.~\eqref{eq:stateW}, we obtain 
\begin{align}
    \rho_{CD}= &\sum_{kl} \gamma_{kl} U_C \left(\rho_1^{(k)}\ox \proj{r}_C\right) U_C^{\dagger} \nonumber \\
    & \qquad\ox U_D \left(\rho_2^{(l)}\ox \proj{r}_D\right) U_D^{\dagger}\nonumber \\
    =& \sum_{kl} \gamma_{kl} \rho_{L_C}^{(k)}\ox \rho_{L_D}^{(l)},
\end{align}
as the joint state of the two Labs after the two Friends completed their measurements. This state is in general entangled, provided that the initial state $\rho_{12}$ was, and can lead to violations of LF inequalities up to the maximal value of $2\sqrt{2}$. This maximum violation occurs if, for example, $A_1= \mathds{1}_1\otimes\proj{0}_C-\mathds{1}_1\otimes\proj{1}_C$, $A_0= \ketbra{0,0}{1,1}_{L_C}-\ketbra{1,1}{0,0}_{L_C}$ and analogously for $B$ (see~\cite{brukner2018no}). As in the QD-E case of Sec.~\ref{ssec:WFwenv}, it is possible to describe a `EWFS-E' case where a decohering environment is included inside Charlie's and Debbie's Labs, $E_C$ and $E_D$ respectively. From the superobservers' points of view this will lead to the following state
\begin{align}
\label{eq:rho_CD_environemt}
    \rho_{CD}=& \sum_{ii'jj'} \bra{i,j}\rho_{1,2} \ket{i',j'} \ketbra{i,c_i,e_i}{i',c_{i'},e_{i'}}_{L_C}\nonumber \\
    & \qquad \ox \ketbra{j,d_j,e_j}{j',d_{j'},e_{j'}}_{L_D},
\end{align}
similar to Eq.~\eqref{eq:stateLab}, where $\ket{c_i}_C$ is the state of Charlie having observed outcome $i$, $\ket{d_j}_D$ that of Debbie having observed outcome $j$ and $\ket{e_k}$ refer to the respective Lab environments recording outcome $k$. If we then consider the original proposal for the measurements by the superobservers, which are on the systems and Friends only, we cannot find violations of the LF-inequalities. As discussed in Sec.~\ref{ssec:The simple Wigner's Friend experiment1} (and similarly in~\cite{20Relano}), any measurements disregarding the environments will give statistics in accordance with the decohered state, 
\begin{align}
\label{eq:rhoCD_deco}
    \rho^{{\rm{Dec}}}_{CD}=& \sum_{ij} \bra{i,j}\rho_{1,2} \ket{i,j} \proj{i,c_i}_{1C}\\
    & \qquad \ox \proj{j,d_j}_{2D},\nonumber
\end{align}
compare Eq.~\eqref{eq:stateF_mixed}. This effective state is separable and can therefore not violate the inequality in Eq.~\eqref{eq:LF-CHSH}. In general, however, Alice and Bob can perform any measurements on their Friends' Labs and, hence,  potentially exploit entanglement present in the state in Eq.~\eqref{eq:rho_CD_environemt}.

\subsection{Quantum Darwinism and Local Friendliness}
\label{ssec:Extended Wigner's Friend experiments2}

\noindent We now discuss EWFS in terms of QD and show how this measurement model severely limits the possibility of observing violations of LF inequalities. We again consider an entangled bipartite state given by Eq.~\eqref{eq:rho12}, where $\rho_1$ is sent to Charlie's isolated Lab $L_C$, and likewise $\rho_2$ to Debbie's Lab $L_D$. For the CHSH-like inequality in Eq.~\eqref{eq:LF-CHSH} Alice and Bob each have two settings, i.e.\ $x,y \in \{0,1\}$, and the setting $1$ is assumed to correspond the the measurement where the superobservers `ask their Friend what they observed' -- analogous to Wigner making an $i$-basis measurement of the Lab in the previous sections. (Likewise setting $0$ would correspond to Wigner in the previous sections making some $j$-basis measurement.)

The first thing to notice is that if, analogously to Sec.~\ref{ssec:WFandQD}, the measurements by Charlie and Debbie are governed by a broadcasting Hamiltonian, LF inequalities cannot be violated. More concretely, consider a Hamiltonian of the form 
\begin{equation}
\begin{aligned}
    &H_{CD}=H_C+H_D\\
    &= \sum_c \ket{c}\bra{c}_1\ox\alpha_C H^{(c)}_C\ox\alpha_{E_C} H^{(c)}_{E_C}\ox\mathbb{1}_{L_D}\\
    &+\sum_d \mathbb{1}_{L_C}\ox\ket{d}\bra{d}_2\ox\alpha_D H^{(d)}_D\ox\alpha_{E_D} H^{(D)}_{E_D},
\end{aligned}
\label{eq:Hamiltonian_EWFS}
\end{equation}
(compare Eq.~\eqref{eq:Hamiltonian_Friend}), from which we obtain an overall post measurement state
\begin{align}
     \rho_{CD}=&\sum_{cd} \bra{c,d}\rho_{12}\ket{c,d}\cdot\left( \proj{c}_1 \bigotimes_{k_C=1}^{N_C}\rho^{\prime(c)}_{k_C}\right)\nonumber \\
    & \qquad \otimes \left( \proj{d}_2 \bigotimes_{k_D=1}^{N_D}\rho^{\prime(d)}_{k_D}\right) \nonumber\\
    %\sum_{klcd} \gamma_{kl} \bra{c,d}\rho^{(k)}_1 \ox \rho^{(l)}_2 \ket{c,d}
    %\left(\proj{c}_1 \ox \rho^{(c)}_{C}\ox \rho^{(c)}_{E_C} %\right)\ox \left(\proj{d}_2 \ox \rho^{(d)}_{D}\ox \rho^{(d)}_{E_D}\right)\nonumber \\
    =& \sum_{cd} p(c,d)\left(\proj{c}_1 \ox \rho^{(c)}_{C}\ox \rho^{(c)}_{E_C} \right)\nonumber\\
    &\qquad \ox \left(\proj{d}_2 \ox \rho^{(d)}_{D}\ox \rho^{(d)}_{E_D}\right)
    \label{eq:rhoCD_inf}
\end{align}
for the two Labs, where $\rho^{(c)}_{C}$ and $\rho^{(d)}_{D}$ are the macrofractions of the Labs $L_C$ and $L_D$ that constitute Charlie and Debbie respectively and $p(c,d)=\bra{c,d}\rho_{12}\ket{c,d}$ is the joint probability for their measured results. Since the state in Eq.~\eqref{eq:rhoCD_inf} is separable, LF-inequalities cannot be violated if the measurement is governed by Eq.~\eqref{eq:Hamiltonian_EWFS}. 

In fact, due to the block-diagonal structure of the Lab states, (resulting from equilibration under broadcasting Hamiltonians, compare Sec.~\ref{ssec:Numerical results1}) there is no coherent superposition between different outcomes corresponding to coherences between the blocks. Hence, the entanglement in the initial system state (due the coherent superpositions of pairs of pointer states) does not lead to entanglement between the two Labs, which is a prerequisite for the violation of LF inequalities.

Hence, while the measurement model considered in this work allows for WF-type effects in the simple WF-QD scenario, for LF-violations in the EWFS-QD case we need to move even further away from SBS states. This means we would need Hamiltonians not of the broadcasting form, which would lead to non-vanishing $\sigma$-like terms, compare Eq.~\eqref{eq:rhoLab1}. The post measurement states would be then of the form 
\begin{align}
\label{eq:CD_generalH}
     \rho_{CD}=&\sum_{cd} q(c,d)\left(\proj{c}_1 \ox \rho^{(c)}_{C}\ox \rho^{(c)}_{E_C} + \sigma_{L_C}\right)\nonumber \\
     & \qquad
     \ox \left(\proj{d}_2 \ox \rho^{(d)}_{D}\ox \rho^{(d)}_{E_D}+\sigma_{L_D}\right).
\end{align}
If the superobservers perform their measurements on this state instead, then there can still be non-classical correlations between the two Labs and hence LF inequality violations.

Moreover, the non-objectivity inherent in our measurement model imposes a fundamental restriction on the potential violations of LF inequalities. More concretely, let us again consider the SBS-like states in Eq.~\eqref{eq:rhoCD_inf}. This post measurement state means that the original LF assumptions in Sec.~\ref{ssec:LF} can no longer be applied. The measurements `revealing what the Friend observed' again are given by the optimal projectors (analogous to the Helstrom measurements) on the Labs of Charlie and Debbie
\begin{align}
    \label{eq:LFx=1}
    A_1=\sum_a a \cdot \mathds{1}_1\ox \Pi^{(a)}_C\ox \mathds{1}_{E_C} \, , \\
    \label{eq:LFy=1}
    B_1=\sum_b b \cdot\mathds{1}_2\ox \Pi^{(b)}_D\ox \mathds{1}_{E_D} \, .
\end{align}
But just as before, these measurements will no longer perfectly reveal the Friends' observed results (compare Eq.~\eqref{eq:WF1}). This modifies the quantities in the AOE assumptions from Sec.~\ref{ssec:LF}. To see this consider for instance
\begin{align}
    \label{eq:reveal_example}
    &P(a|x=1)=\Tr \left( \mathds{1}\ox \Pi^{(a)}_C\ox \mathds{1}_{E_C} \rho_{CD}\right) \\
    &=\sum_{cd} \bra{c,d}\rho_{12}\ket{c,d} \Tr \left( \proj{c}_1 \ox \Pi^{(a)}_C \rho^{(c)}_C \ox \rho^{(c)}_{E_C}\right) \nonumber\\
    &= \sum_c p(c) \Tr \left( \Pi^{(a)}_C \rho^{(c)}_C\right),\nonumber
\end{align}
where we desire that $P(a|c,x=1)=\Tr \left( \Pi^{(a)}_C \rho^{(c)}_C \right)$ is close to $1$ for $a=c$ and close to $0$ for $a\neq c$ (numerically we saw this in the WF-QD case considered earlier). This, however, leads to
\begin{align}
\label{eq:non-idealAOE}
    &P(a=c|x=1)=\sum_c P(c) P(a=c|c,x=1)\\
    &= \sum_c p(c)\Tr \left( \Pi^{(c)}_C \rho^{(c)}_C \right) \geq 1 -\varepsilon, \nonumber
\end{align}
instead of $P(a=c|x=1)=1$ for the ideal EWFS. In our case, if we are considering a two-outcome measurement in the computational basis as in Sec.~\ref{ssec:Numerical results1}, we can say
\begin{equation}
    \label{eq:ourvarepsilon}
    \varepsilon=1- p(0)\Tr\left(\Pi_C^{\left(0\right)}\rho_C^{\left(0\right)}\right)- p(1)\Tr\left(\Pi_C^{\left(1\right)}\rho_C^{\left(1\right)}\right),
\end{equation}
where $p(0)$ and $p(1)$ are the probabilities of each $c$ outcome occurring. In other words, as in the WF-QD case, the `non-idealness' of our result stems from the imperfect fidelity of the $\Pi_C^{\left(c\right)}$ measurements. 

As discussed elsewhere~\cite{moreno2022events,ying2023relating}, for the non-ideal case in Eq.~\eqref{eq:non-idealAOE} we can replace the original AOE-assumption in Sec.~\ref{ssec:LF} by the following weaker versions
\begin{itemize}
    \item{\textbf{AOE$^{\prime}$}\\
    There exists a probability distribution $P(a,b,c,d|x,y)$ such that:
    \begin{itemize}
        \item[1.]{$P(a,b|x,y)=\sum_{c,d}P(a,b,c,d|x,y)$}
        \item[$2^{\prime}$.] {$P(a=c|d,x=1,y)\geq 1-\varepsilon$}
        \item[$3^{\prime}$.]{$P(b=d|c,x,y=1)\geq 1-\varepsilon$}, 
    \end{itemize}  
    }
\end{itemize}
which leads to modified LF-inequalities. For the CSHS-like expression in Eq.~\eqref{eq:LF-CHSH}, these read
\begin{equation}
    \label{eq:LF-CHSH_mod}
    \langle {\rm{CHSH}}\rangle  \leq 2+4\varepsilon,
\end{equation}
which means the larger the non-objectivity is (which in turn means larger values for $\varepsilon$), the harder it is for Alice and Bob to violate the inequality. In fact, once $\varepsilon \geq \frac{\sqrt{2}-1}{2}\approx0.207$, the bound becomes $2\sqrt{2}$ and the modified LF-inequality can no longer be violated in quantum theory. The explanation is that it becomes impossible to discern the true quantum effect (that would, say, exploit contextuality or other quantum resources~\cite{21BaldijaoRobertoWagner}) from an effect with classical origins like measurement imprecision.

It is interesting to compare this to another recent result that studies the emergence of \textit{noncontextuality} in Quantum Darwinism \cite{21BaldijaoRobertoWagner} -- there it is also shown that classicality (there in the form of a noncontextual ontological model) emerges for a sufficiently low probability of failure when performing state discrimination on conditional states of an environment fraction. The $\eta$ in that work is analogous to the $\varepsilon$ here and in \cite{22MorenoNeryDuarte}.

\subsection{Numerical tests of non-objectivity in the EWFS-QD}
\label{ssec:Numerical results2}

\noindent As in the simple WF-QD scenario, we performed numerical simulations of the EWFS with decohering environments. We refer to Appendix~\ref{ssec:EWFSQDnumerics} for details of the model (though much of it follows analogously from Appendices~\ref{ssec:TheMEHBit} and~\ref{ssec:WignersPOVMs}). Since our calculations use only broadcasting Hamiltonians, we were only interested in studying the modifications of LF-inequalities due to non-objectivity, rather than violations of LF-inequalities. Hence, we were only interested in the  equivalents of $P^F(i)$ and $P^W(i)$ and $\varepsilon$, not $P^F(j)$ and $P^W(j)$ for general measurement on the whole Labs. 
In the EWFS-QD case, $P^F(i)$ would correspond to $P^C(c)$ and $P^D(d)$, and $P^W(i)$ would correspond to $P^A(c)$ and $P^B(d)$.
Due to the symmetry of the setup, one can trace out one of the two labs to study the other one individually. In our simulations, we therefore made one Lab as small as possible, $N_D=N_{E_D}=1$, to focus our studies on what happened as $N_C$ and $N_{E_C}$ varied in size.

\begin{figure*}[ht]
    \centering
    \includegraphics[width=0.45\linewidth]{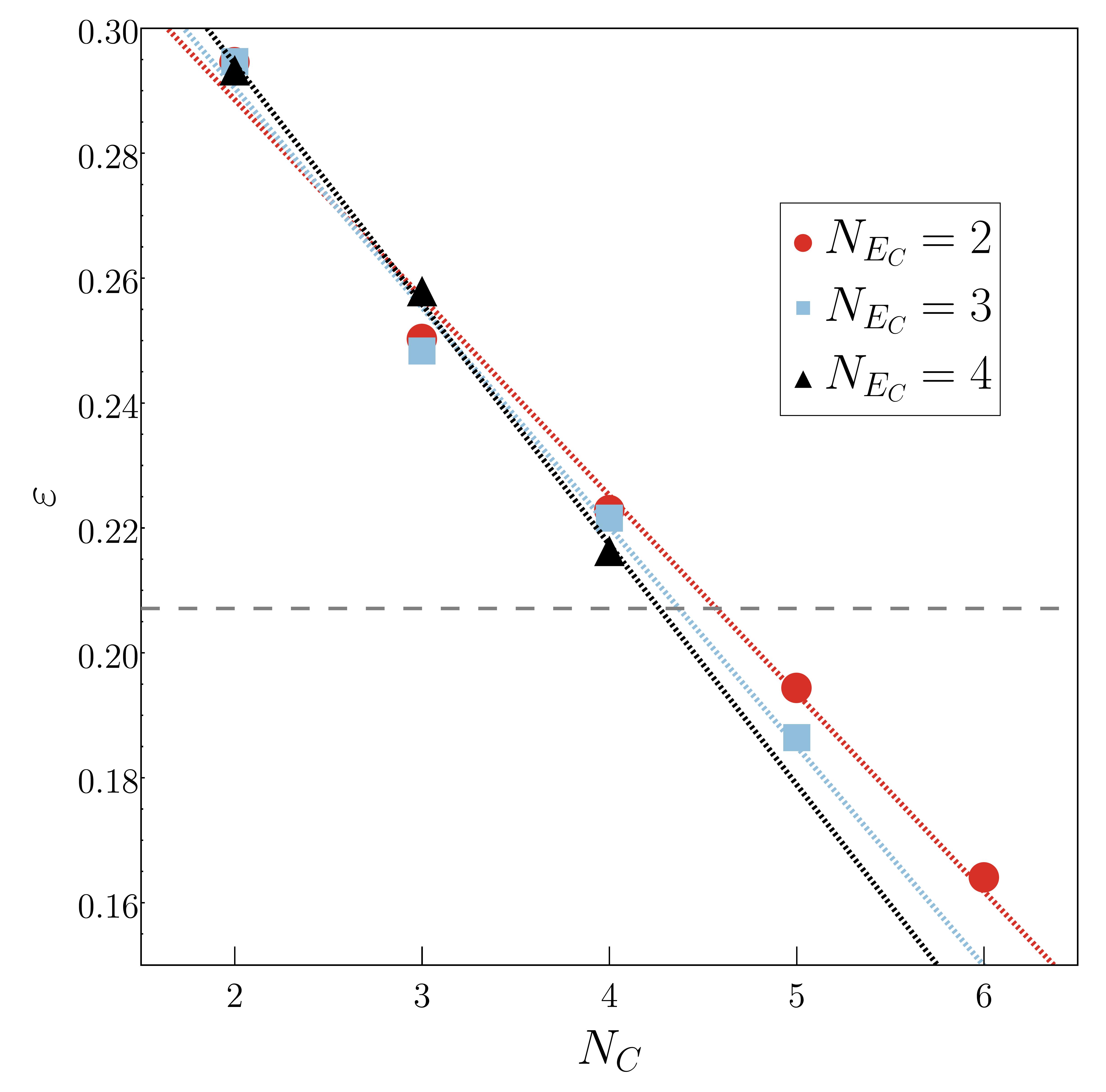}
    \caption{Plot of $\varepsilon$ (see Eq.~\eqref{eq:non-idealAOE}) against $N_{C}$, for different values of $N_{E_C}$ (with $N_D=N_{E_D}=1$ for all points), for $p_0=p_1=0.5$. The dashed line represents $\varepsilon=\left(\sqrt{2}-1\right)/2$, the maximum value below which LF-inequality violations are possible. Plot generated using 200 GUE samples, with negligible SEM. The best-fit lines are solely to guide the eye. }
    \label{fig:PlotEWFSepsilonNE}
\end{figure*}

In particular, we were interested in observing discrepancies $\varepsilon$ between $P^A(c)$ and $P^C(c)$ (the latter probabilities are functionally identical to $p(c)$, the probabilities calculated from quantum theory for the system). In Fig.~\ref{fig:PlotEWFSepsilonNE}, we show how $\varepsilon$ varies with $N_C$, finding that as Charlie grows in size, the non-ideality decreases. This is due to
the non-objectivity decreasing, as in the WF-QD scenario. Notable, for very small values of $N_C$ (2-4 qubits) the values of $\varepsilon$ lie above the $\frac{\sqrt{2}-1}{2}\approx0.207$ -threshold, meaning that LF violations are not possible. Only for \textit{larger} `Charlies' do LF violations become possible. (It should be noted though that for individual instances from the Gaussian Unitary Ensemble that we average over, it may be possible to have values of $\varepsilon$ under the threshold. It is only when taken on average that the LF violations become impossible.)

\section{Conclusions and Outlook}
\label{sec:Conclusions and Outlook}

\noindent In this paper we applied to the Wigner's Friend thought experiment a quantum Darwinism-based measurement model, in order to investigate the emergence of classicality in WF scenarios. Since this model of decoherence contains some non-objectivity \textit{a priori}, it in principle allows for Wigner's-Friend-type effects. These effects are expected to disappear as the Friend and the environment inside her Lab increase in size and the measurement result obtained becomes more and more objective.

When considering the addition of an environment the Friend has only limited access to, and also the inherent non-objectivity in the model, it becomes apparent that disagreement between Wigner and his Friend can stem from a multitude of effects, compare Tab.~\ref{tab:EffectsTable}. For the WF-QD case, this necessitated specifying what we meant by a genuine Wigner's Friend effect. We did this by comparing the disagreement between $W$ and $F$ for the measurement corresponding to Wigner `asking the Friend what she observed' (parameter $\epsilon$) to that for other measurements on the whole Lab (parameter $\Delta$). We asserted that only when $\Delta \gg \epsilon$ could we say that we had WF-type effects. In our simulations we found that we could indeed observe Wigner's Friend effects if both the Friend and the environment were comprised of multiple qubits, with hints that (as expected) we observe an emergence of classicality effect as the Friend increases in size.

Since the broadcasting Hamiltonians to which we limited ourselves permit no mixing between measurement outcomes, all the coherences in the post-measurement state of the Lab are coherences between the Friend and the environment. It is these coherences which Wigner may exploit to produce true WF-type effects (since the Friend does not have access to them), and which are reduced as the number of qubits constituting the Friend (and the environment) increases. This kind of WF-type effect has no direct counterpart in the simplified descriptions of Wigner's Friend experiments where the Lab is modelled as only containing a single qubit for the system and a single qubit for the Friend. There, due to the lack of an environment, any coherences leading to a WF-like discrepancy have to be between the system and the Friend, and would correspond to a `superposition of different measurement outcomes' (compare for example~\cite{vilasini2022general}). We therefore claim that we find a \emph{novel type of Wigner's Friend effect} when adding an environment to the Friend's Lab.
However, when we observe significant disagreement between Wigner and his Friend in Sec.~\ref{ssec:Numerical results1} we also see a substantial Classical Ignorance Error, as estimated by comparing the Bad Friend to Wigner and the regular Friend. Hence, we conclude that the magnitude of this novel WF effect is always comparable to the Classical Ignorance Error. Therefore, a lack of knowledge of coherence information does not hinder a Friend any more than a simple lack of knowledge about the surrounding environment.

Our results might have implications for discussions of what an `agent' means in Wigner's Friend scenarios and in quantum theory more broadly. Often the `Friend' in a WF scenario is assumed to be a simple qubit, but here we show that different effects can be observed as the number of qubits the Friend is composed of changes, even when that number is fairly small. This should be considered in light of discussions of whether the setup needs a Friend that can reason about quantum theory to truly qualify as a WF effect~\cite{nurgalieva2018inadequacy,nurgalieva2022thought,23WisemanCavalcantiRieffel}.

For extended Wigner's Friend setups a `Wigner's Friend effect' is the violation of LF inequalities. Our measurement model severely limits the possibility of observing these. Measurement interactions governed by broadcasting Hamiltonians disallow such violations altogether, since the state of the two Labs after the Friends' measurements is separable. Additionally, the non-objectivity inherent in our model necessitates modifications to the original LF-inequalities, which makes it harder to violate them. Our numerical simulations for this second effect suggest that only larger Friends lead to modifications small enough to still allow for violations of LF inequalities. Provided that there is still an emergence of objectivity if we extend the QD measurement model used in this paper beyond broadcasting Hamiltonians, we find that there are two competing effects at play in EWFS. Like the other deviation from SBS states, the $\sigma_{L_C}$ and $\sigma_{L_D}$ terms in Eq.~\eqref{eq:CD_generalH} (which are necessary for a potential violation of LF inequalities) should decrease for increasing $N_C$ and $N_D$. Hence, the chances of violating an LF inequality due to the two Labs being entangled are highest for \emph{small} Debbie and Charlie. Yet, the required modification of the LF assumptions due to non-objectivity means that \emph{larger} Charlie and Debbie increase the chances of observing an LF violation. Whether there is a range of qubit number where both effects allow for a violation of an LF inequality or whether the QD equilibration measurement model ensures that local Friendliness always holds is an open question which we leave for future work.

\begin{acknowledgements}
We wish to thank Maximilian P.E.\ Lock, Marcus Huber, Nicolai Friis, Florian Meier, Emanuel Schwarzhans, Santiago Bustamante, and Jake Xuereb for assistance and useful discussions. T.R.\ acknowledges support from Templeton Foundation grant No.\ 62423 and Austrian Science Fund (FWF) ESPRIT grant No.\ 7464924. S.E.\ was supported by the UK Engineering and Physical Sciences Research Council grant EP/SO23607/1. V.B.\ is funded by FWF ESPRIT grant No.\ ESP 520-N.
\end{acknowledgements}

\bibliographystyle{unsrt}
\bibliography{refs.bib}
\clearpage

\appendix

\onecolumngrid

\section{Details of the numerical models}
\label{ssec:simple_WF_numerics}

\subsection{The WF-QD dynamical model of decoherence and measurement}
\label{ssec:TheMEHBit}

\noindent In the WF-QD model outlined in the main text, the state $\rho_L$ represents the entire Lab, including the system, the Friend, and the environment, after the Friend has made her measurement. Numerically, this is represented in our model by a broadcasting Hamiltonian being applied to an initially uncorrelated state of S, F, and E. The code to perform this numerical simulation was written in Python using the \textit{QuTiP} package~\cite{12JohanssonNationNori} and other standard Python libraries.

The decoherence process that leads to the SBS-like state, which we refer to as the Friend's measurement, is an uncontrolled unitary evolution governed by a broadcasting Hamiltonian as in~\cite{17KorbiczAguilarCwiklinski,mironowicz2017monitoring,23SchwarzhansBinderHuber}:
\begin{equation} \label{eq:ham}
    H_L=\sum_{i=1}^{d_S} \ket{i}\!\bra{i}_{S} \otimes \sum_{k=1}^{N_{F}}  H_{F_k}^{(i)}\otimes \sum_{k=1}^{N_{E}}  H_{E_k}^{(i)},
\end{equation}
where $\ket{i}$ are the pointer states of the computational basis measurement on the system.
Inspired by random matrix theory, in our simulations we sample the conditional Hamiltonians for each qubit of $F$ and $E$, $H_{F_k}^{(i)}$ and $H_{E_k}^{(i)}$, from the Gaussian Unitary Ensemble (GUE)~\cite{mehta2004random,DAlessio2016,engineer2024equilibration}. Specifically, each conditional Hamiltonian is a matrix of appropriate dimension calculated from $\frac{1}{2}\left(X
+X^{\dagger}\right)$, and the real and imaginary parts of each element of the matrix $X$ are drawn randomly from a standard normal distribution centred on 0 with variance 1. The intention is to make the dynamics of the Friend and environment as generic as possible by making them chaotic. In each of our results shown, a certain number of GUE samples are used and averaged over, to show what a `typical' decoherence process might look like.

As discussed in the main text the broadcasting Hamiltonians eliminate coherences between different pointer states. In the simulations of the WF-QD scenario, we therefore consider the system $S$ to be a single qubit prepared in the state
\begin{equation}    \label{eq:initialS}\rho_S=p_0\ket{0}\!\bra{0}+p_1\ket{1}\!\bra{1},
\end{equation}
where $p_0$ and $p_1$ are the probabilities in the computational basis that the Friend is attempting to extract with her measurement. 
The initial state we study has no coherence in it -- it is merely a classical mixture of outcomes $\ket{0}$ and $\ket{1}$. The quantum effects arise through the system's unitary interaction with an environment under an entangling Hamiltonian. It is the quantum effects arising from the decoherence process itself that Wigner can exploit to produce genuine WF-type effects -- the initial state need not have any quantum properties to begin with.

The Friend and the environment within the Lab are represented by $N_F$ and $N_E$ qubits respectively. Both $F$ and $E$ are separately coupled to the system $S$, but crucially not to each other. All of $F$ and $E$'s qubits are initialised in the $\ket{0}$ state, so the total initial state is 
\begin{equation}    \label{eq:initialprep}\rho_L^{{\rm{init}}}=\left(p_0\ket{0}\!\bra{0}_S+p_1\ket{1}\!\bra{1}_S\right)\bigotimes_{k=1}^{N}\ket{0}\!\bra{0}_k,
\end{equation}
where $N=N_F+N_E$. After the decoherence process under the Hamiltonian $H_L$ has finished, the state of the Lab can be represented by Eq.~\eqref{eq:rhoLab2} in the main text (with $0<{\rm{Tr}}\left(\rho_F^{\left(i_1\right)}\rho_F^{\left(i_2\right)}\right)\ll 1\,\, \forall\,\,\,i_1,\,\,i_2$ and equivalent for $\rho_E^{\left(i\right)}$). The assumption of the environments all being initialised in the $\ket{0}$ state is quite restrictive and non-natural, but suffices for the illustrative numerical examples we provide here.

\subsection{Wigner's Measurements}
\label{ssec:WignersPOVMs}

\noindent In the WF-QD scenario, Wigner chooses to make one of two different measurements: an $i$-basis measurement on the Friend's quantum system $\rho_F$, or a $j$-basis measurement on the Lab as a whole. In conventional WF scenarios, it is assumed that the former is trivial. However here we adopt the quantum Darwinism perspective that even this measurement is non-trivial to perform. For the two-outcome measurements considered here, the optimal measurement that Wigner can perform to maximise his chance of successfully discerning between $\rho_F^{\left(i=0\right)}$ and $\rho_F^{\left(i=1\right)}$ is the Helstrom measurement~\cite{09BarnettCroke}, a POVM on $\mathcal{H}_F$ with elements $\{\Pi_F^0,\Pi_F^1\}$. We begin by defining $\hat{O}_F=p_0\rho_F^0-p_1\rho_F^1$, an operator derived from the states $\rho_F^{\left(i\right)}$. We find its eigenvalues and eigenvectors $\{\lambda_{l},\Pi^O_{l}\}_{l}$, and from that we define
\begin{equation}
\Pi_F^0 = \sum_{l:\lambda_{l} < 0} \Pi^O_{l}, \qquad
\Pi_F^1 = \sum_{l:\lambda_{l} > 0} \Pi^O_{l}.
\label{eq:HelstromM}
\end{equation}
(If any $\lambda_{l}$ happen to be identically zero they can be arbitrarily assigned to either POVM element.) It is known that this measurement minimises the chances of State Discrimination Errors. In principle, it is possible for Wigner to also ask the \textit{environment} what it saw, and so we also defined and calculated equivalent Helstrom POVM elements on $\mathcal{H}_E$ using $\{\rho_E^{\left(0\right)}, \rho_E^{\left(1\right)}\}$: $\{\Pi_E^0,\Pi_E^1\}$.

Then Wigner's measurement of the Friend -- him `asking her what she saw' -- is given by:
\begin{equation}
    \label{eq:WhatSheSaw}
    P^W(i)=\Tr \left( \left(\mathds{1}_S\otimes \Pi_F^{(i)}\otimes\mathds{1}_E\right)\rho_L\right).
\end{equation}

(One could argue that we are employing circular reasoning by using the states $\rho_F^{\left(i\right)}$ to construct this POVM, but we also implemented a `maximally discerning POVM' numerically using a Python implementation of CVXPY~\cite{16DiamondBoyd}, inspired by methods from~\cite{engineer2024equilibration}. We found that it always almost perfectly coincided with the Helstrom measurement.)

We then constructed the $j$-basis measurement on the entire $\mathcal{H}_L$ from combinations of the Helstrom POVM elements $\{\Pi_F^0,\Pi_F^1\}$ and $\{\Pi_E^0,\Pi_E^1\}$, plus `POVM elements' on the system: $\{\Pi_S^0=\ket{0}\!\bra{0}_S,\Pi_S^1=\ket{1}\!\bra{1}_S\}$. First, we constructed an eight-element POVM from all possible combinations of the POVM elements on $S$, $F$, and $E$: $M_L^{\alpha\beta\gamma}= \Pi_S^{\alpha}\otimes \Pi_F^{\beta}\otimes \Pi_E^{\gamma}$. Then from these POVM elements, we constructed the following two-outcome POVM:
\begin{equation}
    \label{eq:sameordiffpovms}
    \begin{aligned}
      M_L^{0}&= \Pi_S^0\otimes \Pi_F^0\otimes \Pi_E^0 + \Pi_S^0\otimes \Pi_F^1\otimes \Pi_E^1 + \Pi_S^1\otimes \Pi_F^0\otimes \Pi_E^0 + \Pi_S^1\otimes \Pi_F^1\otimes \Pi_E^1\\
      M_L^{1}&= \Pi_S^0\otimes \Pi_F^0\otimes \Pi_E^1 + \Pi_S^0\otimes \Pi_F^1\otimes \Pi_E^0 + \Pi_S^1\otimes \Pi_F^0\otimes \Pi_E^1 + \Pi_S^1\otimes \Pi_F^1\otimes \Pi_E^0.
    \end{aligned}
\end{equation}
This POVM essentially ask the question: are the indices of $F$ and $E$ the same or different? $M_L^0$ corresponds to `same', and $M_L^1$ to `different'. The POVM checks whether the Friend and the environment observe the same outcome. This is a key requirement of \textit{objectivity}, a central concern in the study of quantum Darwinism. It is this two-outcome POVM which we use as the $j$-basis measurement in the numerical studies of this work, to search for differences between Wigner and the Friend's predictions for measurement outcomes.

\subsection{Details of the EWFS-QD model}
\label{ssec:EWFSQDnumerics}

\noindent Much of the numerical implementation of the EWFS-QD model is identical to that of the WF-QD scenario's implementation. Charlie and Debbie each have a qubit that is half of an entangled pair, and they perform a two-outcome measurement in the computational basis on their qubit. Each of them are made of a number of qubits $N_C$ and $N_D$, and each are surrounded by an environment made of a number of qubits $N_{E_C}$ and $N_{E_D}$. The initial state of the entangled pair is $\rho_{12}=\ket{\psi_{12}}\!\bra{\psi_{12}}$, with $\ket{\psi_{12}}$ given by Eq.~\eqref{eq:psi12} in the main text. The initial state of the two labs $L_C$ and $L_D$ combined is:
\begin{equation}    \label{eq:initialprepCD}\rho_{CD}^{{\rm{init}}}=\rho_{12}\bigotimes_{k=1}^{N}\ket{0}\!\bra{0}_k,
\end{equation}
where now we have $N=N_C+N_D+N_{E_C}+N_{E_D}$. (The choice of $\theta$ in $\ket{\psi_{12}}$ is irrelevant.)

For the dynamics, we use the EWFS equivalent of Eq.~\eqref{eq:ham} (compare to Eq.~\eqref{eq:Hamiltonian_EWFS} in the main text):
\begin{equation}
\begin{aligned}
    H_{CD}&=H_C+H_D\\
    &= \sum_c \ket{c}\bra{c}_1\ox H^{(c)}_C\ox H^{(c)}_{E_C}\ox\mathbb{1}_2\ox\mathbb{1}_D\ox\mathbb{1}_{E_D}\\
    &+\sum_d \mathbb{1}_1\ox\mathbb{1}_C\ox\mathbb{1}_{E_C}\ox\ket{d}\bra{d}_2\ox H^{(d)}_D\ox H^{(D)}_{E_D},
\end{aligned}
\label{eq:Hamiltonian_EWFS_GUE}
\end{equation}
where each of the $H^{(c)}_C$, $H^{(c)}_{E_C}$, $H^{(d)}_D$, and $H^{(d)}_{E_D}$ are separately sampled from the GUE in the same way as before. The post-measurement state of the two labs combined, $\rho_{CD}$, is calculated as before using the pinching map method. 
The measurements that the superobservers Alice and Bob performe were once again derived from Helstrom measurements, and their POVM elements have analogous forms to Eq.~\eqref{eq:HelstromM}.

\section{Superobservers' measurements in terms of quantum Darwinism}
\label{app:WmsmQD}

\noindent Naturally, given the discussion of modelling the Friend's measurement as a decohering equilibration to an SBS-like state, one may ask if the same description can be applied to Wigner's measurement of the Lab as a whole. (There is possible justification for considering a chain of superobservers in the EWFS, see for instance~\cite{szangolies2020quantum}) To do so, we imagine Wigner is a collection of $N_W$ qubits in a Lab $L_W$, and his Lab also contains both the Friend's Lab (effectively the `system' being measured), and another collection of $N_{E_W}$ qubits representing Wigner's environment. In the WF-QD scenario, for each outcome $j$ on the Friend's Lab, we label the commensurate states of $W$ or $E_W$ as $\sigma_{k'}^{\left(j\right)}$. From the perspective of a hypothetical super-superobserver(!), the post-measurement state of Wigner's Lab is in the Hilbert space $\mathcal{H}_{L_W}=\mathcal{H}_L\ox\mathcal{H}_{W}\ox\mathcal{H}_{E_W}$, and has the form: 
\begin{equation}
    \label{eq:rhoW}
    \rho_{{L_W}}=\sum_{j}P_j \proj{j}_{L} \bigotimes_{k'=1}^{M} \sigma_{k'}^{\left(j\right)}= \sum_{j}P_j \proj{j}_{L} \otimes \sigma^{(j)}_W \otimes \sigma^{(j)}_{E_W}.
\end{equation}
As before, $P(j) = \Tr(\proj{j}\rho_L)=P^W(j)$ is the probability of Wigner observing outcome $j$ for a measurement of $\rho_L$ given by Eq.~\eqref{eq:rhoLab2} in the main text. If we were to consider further measurements on Wigner by the super-superobserver we would then need to compare $P^W(j)$ with this super-superobserver's predictions for W's $j$ outcomes. But crucially, none of this impacts the model described in the main text. For the WF-QD case, we need only the state of the Friend's Lab $\rho_L$ for calculating the relevant probabilities. Hence,  we do not need an explicit model of Wigner's Lab.

Likewise in the EWFS-QD model, from the perspective of super-superobservers potentially measuring Alice and Bob, the measurements by Alice and Bob can be described as equilibration processes towards SBS-like states, giving 
\begin{align}
    \rho_{AB|XY}&=\sum_{abcd} \bra{c,d}\rho_{12} \ket{c,d}  \bra{a_x}\rho_{L_C}^{(c)} \ket{a_x} 
    \otimes \bra{b_y} \rho_{L_D}^{(d)} \ket{b_y} \nonumber\\
    & \quad  \otimes
    \left( \proj{a_x}_{L_C} 
    \bigotimes_{k_A=1}^{M_A}\sigma^{(a)}_{k_A}\right) \ox \left(\proj{b_y}_{L_D} \bigotimes_{k_B=1}^{M_B}\sigma^{(b)}_{k_B}\right)
    \nonumber \\
    &=\sum_{ab} P(a,b|x,y) \proj{a_x}_{L_C} \otimes \sigma^{(a)}_{A} \ox \sigma^{(a)}_{E_A} \ox \proj{b_y}_{L_D} \ox \sigma^{(b)}_{B}\ox \sigma^{(b)}_{E_B},
    \label{eq:rhoAB_inf}
\end{align}
where we defined $\rho^{(c)}_{L_C}= \proj{c}_1\ox \rho^{(c)}_C\ox\rho^{(c)}_{E_C}$ and accordingly $\rho^{(d)}_{L_D}$. 
We also define $\sigma_{k_{A}}^{(a)}$ as the states of the $N_A$ qubits representing Alice and the $N_{E_A}$ qubits representing the environment surrounding Alice, each commensurate with having observed outcome $a$ (and analogously $\sigma_{k_{B}}^{(b)}$ for Bob and his environment). Alice and Bob's measurements are described by projectors $M_{(a|x)}=\proj{a_x}$ and $M_{(b|y)}=\proj{b_y}$ which depend on their inputs $x,y$. The random numbers $x,y$ determine sets of POVMs $\{M_{(m|0)}\}$, $\{M_{(m|1)} \}$, etc.\ for each superobserver, i.e.\ $m=a,b$. Therefore, the post measurement states are conditional on which of these measurements Alice and Bob perform. 

Note that in~\cite{20Relano}, measurements by the superobservers are included in the EWFS considered there. But when calculating the overall state, the fact that they perform different measurements depending on their inputs is not taken into account. For the initial state and observables from~\cite{brukner2018no}, a value of $\langle \rm{CHSH} \rangle= 1/\sqrt{2}$ is derived, implying that the model does not permit CHSH inequality violation, but we argue that this results from using only one out of four conditional post measurement states for the superobservers.

In the context of LF inequalities, all we are interested in are the probabilities
\begin{equation}
  P(a,b|x,y)= \bra{a_x,b_y} \left(\sum_{cd} \bra{c,d}\rho_{12} \ket{c,d}  \rho_{L_C}^{(c)} 
    \otimes \rho_{L_D}^{(d)} \right) \ket{a_x,b_y}
    = \Tr \left( M_{(a|x)} \ox M_{(b|y)} \rho_{CD} \right).
\end{equation}
Clearly, calculating these probabilities in terms of the decoherence model of~\cite{20Relano} or in terms of the QD-based model of this work only requires a dynamical description of Charlie's and Debbie's measurements in order to obtain $\rho_{CD}$ (see Eq.\eqref{eq:rhoCD_inf} in the main text). Hence, the problem of having to use conditional states $\rho_{AB|XY}$ can be avoided entirely and we do not explicitly model Alice and Bob's measurements outside of in this appendix. 

\section{Additional numerical results for WF-QD}
\label{sec:puneven}

\begin{figure}[h!]
    \centering
    \includegraphics[width=0.5\linewidth]{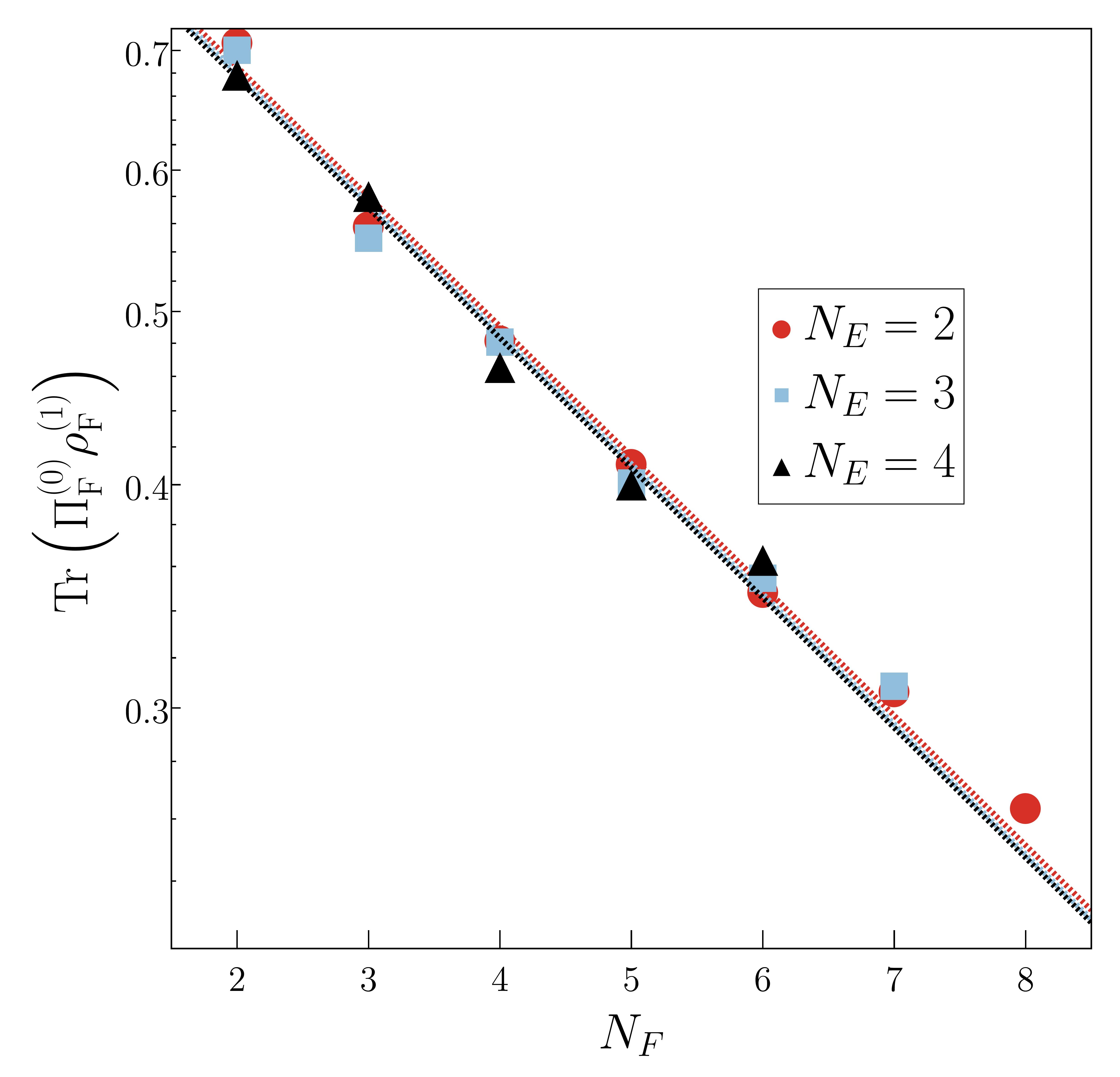}
    \caption{Plot (on a log-scale) of the non-objectivity in the Friend, as measured by the overlap ${\rm{Tr}}\left(\Pi_F^0\rho_F^{1}\right)$, as a function of $N_F$, for different values of $N_E$, when $p_0=075$ and $p_1=0.25$. Note this is different from ${\rm{Tr}}\left(\Pi_F^1\rho_F^{0}\right)$ used in Fig.~\ref{fig:OverlapsDecreasing} in the main text. 200 GUE samples were used to generate the points in these plots, with negligible SEM. The logarithmic best-fit lines are solely to guide the eye.}
    \label{fig:PlotM0F1vsNFune}
\end{figure}

\noindent Here we present additional numerical results for the WF-QD case when $p_0\neq p_1$, specifically when $p_0=0.75$. First, in Fig.~\ref{fig:PlotM0F1vsNFune}, we see a plot of ${\rm{Tr}}\left(\Pi_F^0\rho_F^{1}\right)$, which, like the equivalent Fig.~\ref{fig:OverlapsDecreasing} in the main text, substantially decreases with $N_F$. One key difference, however, is that the plot in Fig.~\ref{fig:PlotM0F1vsNFune} starts and ends at much larger values than that in  Fig.~\ref{fig:OverlapsDecreasing}, indicating that the uneven probability is affecting how well the $\Pi_F^1$ Helstrom measurement is working. Note also that here we plot ${\rm{Tr}}\left(\Pi_F^0\rho_F^{1}\right)$, as opposed to the ${\rm{Tr}}\left(\Pi_F^1\rho_F^{0}\right)$ that we plot in Fig.~\ref{fig:OverlapsDecreasing} (in the even-probability case the two are almost identical). In the uneven case, due to the nature of the Helstrom measurement, the measurement ${\rm{Tr}}\left(\Pi_F^1\rho_F^{0}\right)$ is always close to 0 (less than 0.1 in all uneven cases studied here). This is because it is always more efficient to guess you are in the higher probability state. In the search for the emergence of objectivity that we pursue here, the interesting behaviour is thus in ${\rm{Tr}}\left(\Pi_F^0\rho_F^{1}\right)$, which we show decreases rapidly as $N_F$ increases.

\begin{figure}[h!]
    \centering    
    \includegraphics[width=0.5\linewidth]{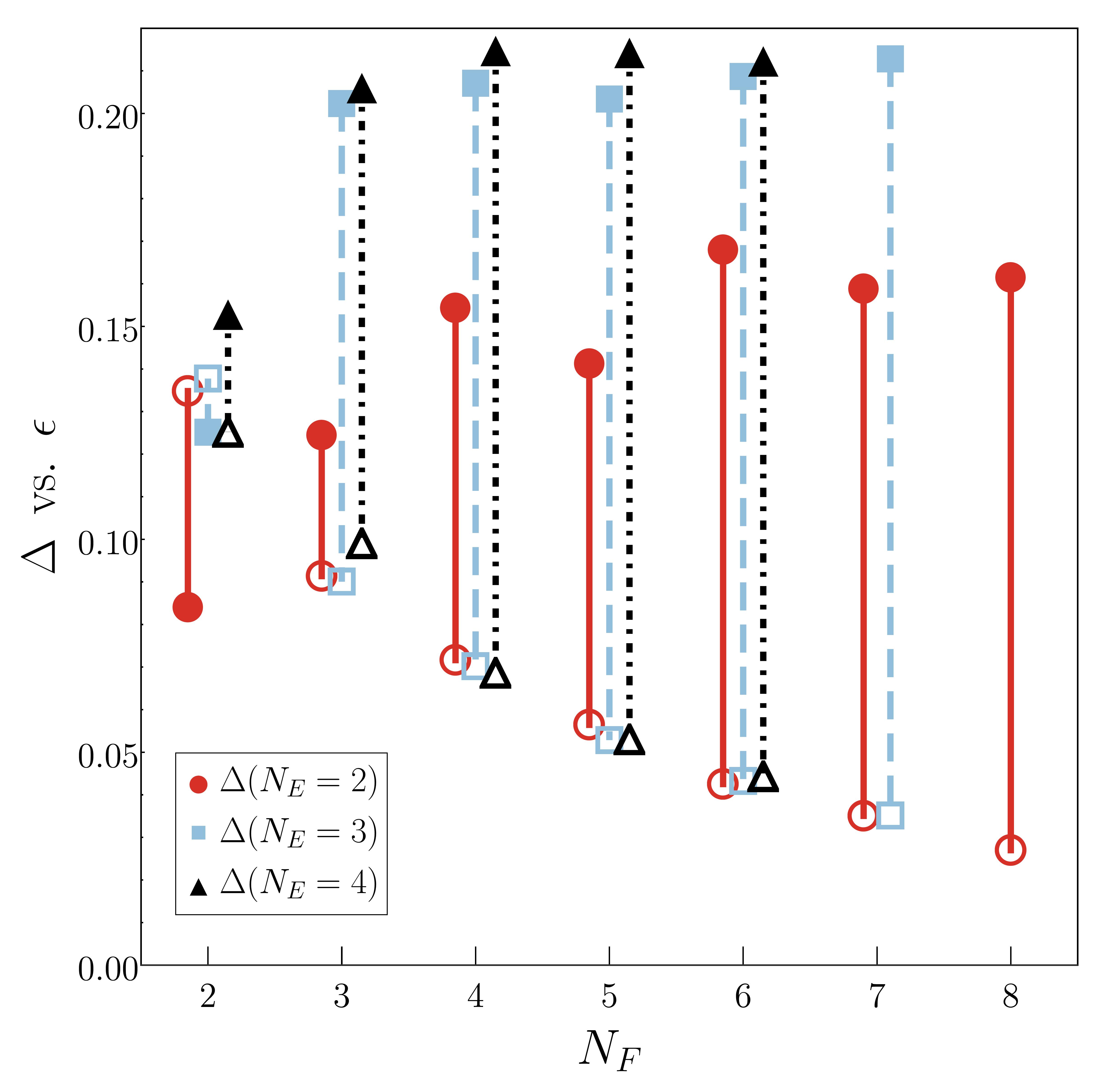}
    \caption{Plot of the difference $\Delta=|P^W(j=0)-P^F(j=0)|$ (solid data points), compared to the difference $\epsilon=|P^W(i=0)-P^F(i=0)|$ (hollow data points). $\Delta$ and $\epsilon$ vary with $N_F$, for different values of $N_E$, when $p_0=0.75$ and $p_1=0.25$. Plot generated using 200 GUE samples, with negligible SEM. $\Delta$ compares Wigner's measurement outcome $P^W(j=0)$ to the Friend's prediction of Wigner's outcome $P^F(j=0)$, and $\epsilon$ is the difference between the Friend's measurement outcome $P^F(i=0)$ and Wigner's measurement outcome $P^W(i=0)$.}
    \label{fig:Deltavsepsilonune}
\end{figure}

When comparing again $\Delta=|P^W(j=0)-P^F(j=0)|$  and $\epsilon=|P^W(i=0)-P^F(i=0)|$ as functions of $N_F$ for different values of $N_E$, here we observe significantly different behaviour for $p_0\neq p_1$ as compared to the $p_0=p_1$ case in the main text -- compare Fig.~\ref{fig:Deltavsepsilonune} to Fig.~\ref{Fig:Plotepsilondelta}. Here, surprisingly, we find that $\Delta$ does not decrease as $N_F$ and $N_E$ increase (with possible weak evidence that it \textit{increases} -- the SEM remains low for each data point but the are insufficient data points to recognise a clear trend). This is in direct opposition to the result in the main text, where we argued that it may be approaching $P^F(j=0)$ and demonstrating the emergence of classicality. This is partly explained by a change in the behaviour of $P^F(j=0)$, since here it is not always equal to 1. It never drops below approximately 0.9 in the cases we study, however, so it cannot be the primary source of the deviation.

More important in explaining this change in behaviour is the fact that $\epsilon$ is much larger in Fig.~\ref{fig:Deltavsepsilonune} than in Fig.~\ref{Fig:Plotepsilondelta}. We see $P^W(i=0)$ differing significantly from 0.75 for all $N_F$, $N_E$ ($P^F(i=0)$ still remains negligibly different from 0.75). One possible explanation for this is that the non-idealness of Wigner's measurement, ${\rm{Tr}}\left(\Pi_F^0\rho_F^{1}\right)$, varies much more strongly as a function of $N_F$ than the small `emergence of classicality effect' hinted at in Fig.~\ref{Fig:Plotepsilondelta}. The former effect can serve to increase $\Delta$, since it improves the extent to which Wigner's $j$-basis measurement actually measures whether F and E agree with each other. This means that it is possible that in the unequal probability case, the decrease in non-idealness `wins out' over the emergence of classicality effect.

For $p_0=p_1$ in the main text we asserted that we observe a genuine Wigner's Friend effect, since $\Delta$ was clearly much larger than $\epsilon$ throughout the simulations. For the case of $p_0=0.75 \neq p_1=0.25$ depicted in Fig.~\ref{fig:Deltavsepsilonune}, the situation is clearly different. By visual inspection we conclude that there is no genuine Wigner's friend effect for the specific $j$-measurement we considered. While for some of the data points we find that $\Delta>\epsilon$, unlike in the equal probability case, we do not clearly have $\Delta\gg\epsilon$ for any of them. In fact, when $N_E=2$, the points lie within each others' SEM (even though this quantity is small for each data point and we do not plot it), so we cannot even claim to have $\Delta>\epsilon$.

\begin{figure}[h!]
    \centering
    \includegraphics[width=0.45\linewidth]{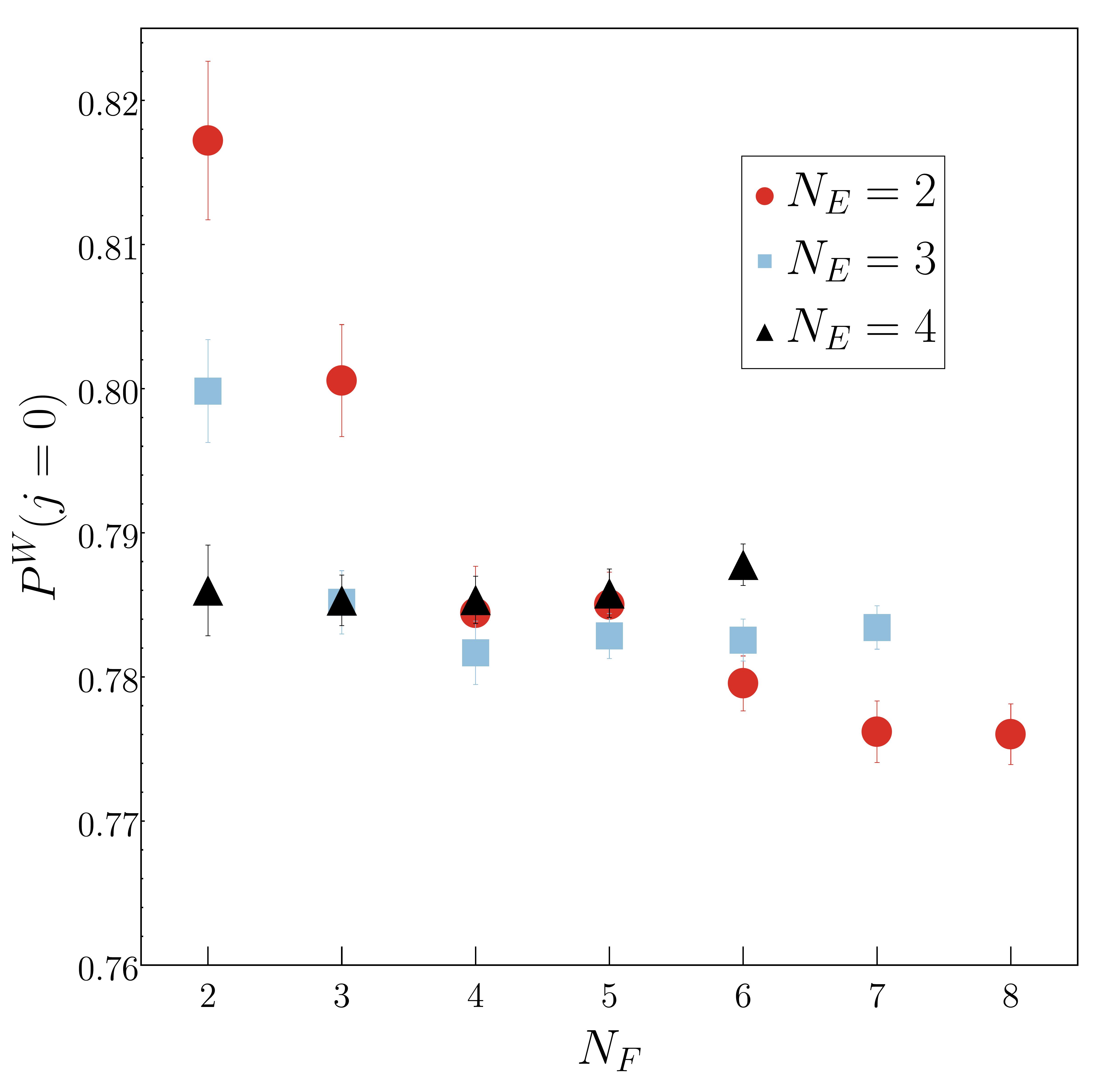}
    \includegraphics[width=0.45\linewidth]{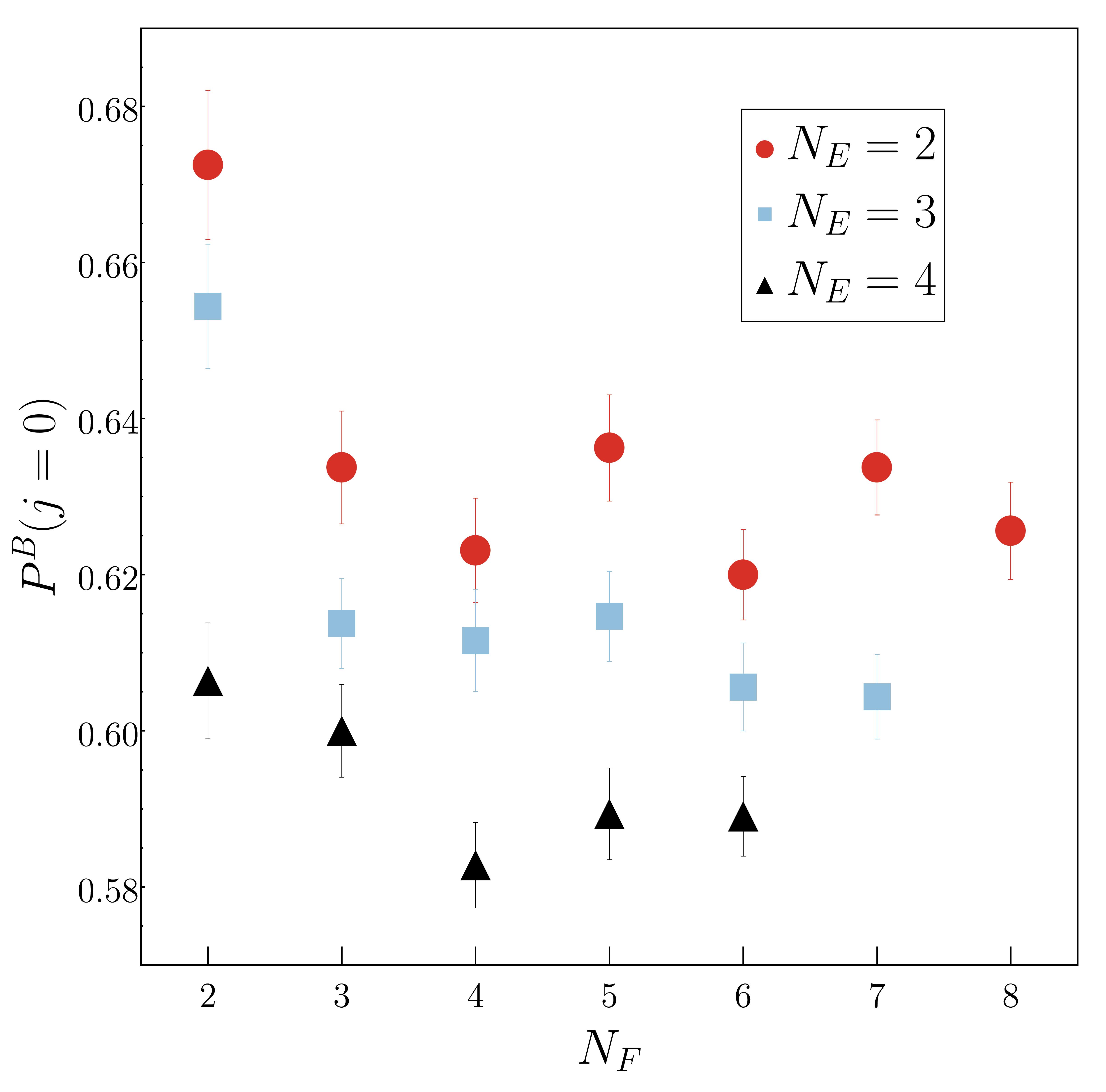}
    \caption{\textit{Left:} Plot of Wigner's measurement outcome $P^W(j=0)$ when using the POVM with elements $\{M_L^0,M_L^1\}$ (see Appendix~\ref{ssec:WignersPOVMs}), against $N_F$, for different values of $N_E$, when $p_0=0.75$ and $p_1=0.25$. Plot generated using 200 GUE samples. Unlike the other plots in this work, here the error bars representing the SEM are non-negligible. \textit{Right:} Plot of the Bad Friend's predictions $P^B(j=0)$ in the same scenario. In the even-probability case, this was always identically 0.5, but here we find it varies with $N_F$ and $N_E$. (SEM error bars included.))}
    \label{fig:PBWjsune}
\end{figure}

Lastly, we also plot in Fig.~\ref{fig:PBWjsune} the value of $P^W(j=0)$ itself used in the calculation of $\Delta$ in Fig.~\ref{fig:Deltavsepsilonune}. This quantity is plotted as a function of $N_E$ for different values of $N_F$. In the uneven probability case considered here it clearly does not tend towards agreement with $P^F(j=0)$ as $N_F$ increases, as happened in the even-probability case (for the reasons outlined above). In this same figure we also show $P^B(j=0)$, the Bad Friend's predictions for Wigner's measurement outcomes, for this same scenario. In the main text we did not include a plot of this since it was always almost exactly 0.5 regardless of $N_E$ and $N_F$. Here it is still always substantially different from $P^F(j=0)$, as $P^B(j=0)$ varies from approximately 0.55 to approximately 0.7, whereas $P^F(j=0)$ is never less than approximately 0.9 and is usually close to 1. In Fig.~\ref{fig:PBWjsune}, the SEM for some of the data points is sufficiently large that we find value in plotting it, though this does not affect anything qualitative about the analysis.

\end{document}